\documentclass[aps,prd,floatfix,12pt, notitlepage,a4paper,floatfix,showpacs,nofootinbib,superscriptaddress,tikz]{revtex4-2}
\pdfoutput=1
\RequirePackage[colorlinks=true,urlcolor=red,anchorcolor=red,citecolor=red,filecolor=red,linkcolor=red,menucolor=red,pagecolor=red,linktocpage=true,pdfproducer=medialab,pdfa=true]{hyperref}
\usepackage[usenames,dvipsnames]{color}  
\usepackage{graphicx}
\usepackage{comment}
\usepackage{xcolor}
\usepackage{bm}
\usepackage{bbold}
\usepackage{latexsym}
\usepackage{amsthm}
\usepackage{upgreek}
\usepackage{amsmath}
\usepackage{placeins}
\usepackage{wrapfig}
\usepackage{amssymb}
\usepackage{cancel}
\usepackage{multirow}
\usepackage{rotating}
\usepackage{geometry}
\usepackage{slashed}
\usepackage{hhline}
\usepackage{amsmath}
\usepackage{fancybox}
\usepackage[utf8]{inputenc}
\usepackage[normalem]{ulem} 
\usepackage[normalem]{ulem}
\usepackage{color,soul}
\setul{0.5ex}{0.3ex}
\setulcolor{blue}
\usepackage{braket}
\definecolor{linkcolor}{rgb}{0,0,0.5}
\allowdisplaybreaks
\allowdisplaybreaks[1]
\allowdisplaybreaks[2]

\definecolor{greenLinks}{rgb}{0, 0.6, 0}
\definecolor{blueLinks}{rgb}{0, 0, 0.6}
\definecolor{redLinks}{rgb}{0.6, 0, 0}
\definecolor{tempText}{rgb}{0.55, 0.10,0.67}
\definecolor{eprintLinks}{rgb}{0.4, 0.4, 0.4}
\definecolor{journalLinks}{rgb}{0.6, 0, 0}
\newcommand {\ignore}[1]{}
\usepackage{soul}
\definecolor{mightnightblue}{RGB}{25,25,112}
\definecolor{brown}{rgb}{0.59, 0.29, 0.0}
\definecolor{darkred}{rgb}{0.6,0,0}

\def\SM{$\mathrm{ SU(3)_C \otimes SU(2)_L \otimes U(1)_Y }$ }

\def\lsim{\mathrel{\rlap{\lower4pt\hbox{\hskip1pt$\sim$}}
    \raise1pt\hbox{$<$}}}
\def\gsim{\mathrel{\rlap{\lower4pt\hbox{\hskip1pt$\sim$}}
    \raise1pt\hbox{$>$}}}

\newcommand{\AddrMPI}{Max-Planck-Institut f\"ur Kernphysik, Saupfercheckweg 1, 69117 Heidelberg, GERMANY}
 
\newcommand{\AddrBhopal}{Department of Physics, Indian Institute of Science Education and Research - Bhopal, Bhopal Bypass Road, Bhauri, Bhopal 462066, INDIA}

\begin{document}
\title{Comprehensive Phenomenology of the Dirac Scotogenic \\ Model:
Novel Low-Mass Dark Matter
}

\author{Salvador Centelles Chuli\'{a}}\email{chulia@mpi-hd.mpg.de}
\affiliation{\AddrMPI}
\author{Rahul Srivastava}\email{rahul@iiserb.ac.in}
\affiliation{\AddrBhopal}
\author{Sushant Yadav}\email{sushant20@iiserb.ac.in}
\affiliation{\AddrBhopal}

\begin{abstract}
\vspace{1 cm}
\noindent
The Dirac Scotogenic model provides an elegant mechanism for generating small Dirac neutrino masses at the one-loop level.  A single abelian discrete $\mathcal{Z}_6$ symmetry simultaneously protects the ``Diracness'' of the neutrinos and the stability of the dark matter candidate. This symmetry originates as an unbroken subgroup of the so-called \textit{445} $U(1)_{B-L}$ symmetry.
 Here, we thoroughly explore the phenomenological implications of this construction, including an analysis of 
 electroweak  vacuum stability, charged lepton flavor violation, and the dark matter phenomenology. After considering all constraints, we also show that the model allows for the possibility of novel low-mass scalar and fermionic dark matter, a feature not shared by its canonical Majorana counterpart. 
 
\end{abstract}

\maketitle 

\tableofcontents

\section{Introduction}
The detection of neutrino oscillations \cite{Super-Kamiokande:1998kpq} implying the existence of massive neutrinos, the observation that approximately 27 \% of the total mass-energy content in the Universe is dark matter (DM) \cite{Bertone:2004pz, Planck:2018vyg} and the baryon asymmetry of the universe serve as compelling indications of the existence of Beyond Standard Model (BSM) physics. A notable proposal in this context is the ``Scotogenic Model", in which neutrino masses are generated through a one-loop process involving the DM candidates \cite{Tao:1996vb,Ma:2006km}. In its canonical form, the model relies on the introduction of an `\textit{ad hoc}' $\mathcal{Z}_2$ symmetry to guarantee the stability of the DM. Although conventional neutrino mass models, including the original scotogenic model, typically predict Majorana light neutrinos, experimental evidence supporting their Majorana nature such as neutrinoless double beta decay \cite{KamLAND-Zen:2016pfg} remains elusive. Consequently, there has been growing interest in exploring the origins of light Dirac neutrino masses at sub-eV scales \cite{Ma:2014qra,Ma:2015raa,Ma:2015mjd,Bonilla:2016diq,CentellesChulia:2016fxr,CentellesChulia:2017koy,Bonilla:2018ynb,CentellesChulia:2018gwr,CentellesChulia:2018bkz,CentellesChulia:2019xky,Peinado:2019mrn,Wang:2017mcy,Borah:2017leo,Hirsch:2017col,Jana:2019mgj,Jana:2019mez,Calle:2019mxn,Nanda:2019nqy,Ma:2019byo,Ma:2019iwj,Correia:2019vbn,Saad:2019bqf,Ma:2019yfo,CentellesChulia:2020dfh,Luo:2020sho,Luo:2020fdt,CentellesChulia:2020bnf,Guo:2020qin,delaVega:2020jcp,Borgohain:2020csn,Leite:2020wjl,Chulia:2021jgv,Bernal:2021ppq,Biswas:2021kio,Mahanta:2021plx,Hazarika:2022tlc,CentellesChulia:2022vpz,Maharathy:2022gki,Berbig:2022pye,Mishra:2021ilq,Chowdhury:2022jde,Biswas:2022vkq,Li:2022chc,Correia:2019vbn,Berbig:2023uzs,Mahapatra:2023oyh,Borah:2024gql,Singh:2024imk}, formulated either at the tree or loop level, as well as their connections with different physics sectors and their phenomenology. While the canonical scotogenic model was designed for Majorana neutrinos, the scotogenic framework is general enough to be applicable to Dirac neutrinos both at one-loop level \cite{Bonilla:2018ynb, Guo:2020qin,Hazarika:2022tlc,CentellesChulia:2022vpz} as well as at two loop level \cite{Bonilla:2016diq,CentellesChulia:2019xky}. 

In this discussion, we focus on comprehensively exploring the phenomenology of a simplified version of the Dirac Scotogenic model proposed in \cite{Bonilla:2018ynb}. The chiral structure of $U(1)_{B-L}$, under which the three right-handed neutrinos transform as $\nu_R \sim (-4, -4, 5)$\cite{Montero:2007cd,Ma:2014qra}, does not allow the tree level Dirac mass term $\bar{L} \tilde{H} \nu_R$ and forbids the effective Majorana neutrino mass operators at all orders. It is also anomaly free and leads to one massless neutrino. Here, we explicitly and softly break $U(1)_{B-L} \to \mathcal{Z}_6$, preserving the Dirac nature of neutrinos but allowing a one-loop realization of the operator $\bar{L} \tilde{H} \chi \nu_R$. The particles running inside the loop (a scalar gauge singlet $\xi$, a scalar $SU(2)_L$ doublet $\eta$ and a neutral fermion vector-like pair $N_L, \, N_R$) form the `dark sector', the lightest of them being the DM candidate and its stability protected by the remnant $\mathcal{Z}_6$ symmetry. While the canonical Majorana scotogenic model can only have a doublet scalar or a neutral fermion as DM, this framework allows the additional option of a singlet scalar DM.
 
This paper is organized as follows. Sec.~\ref{sec:model} provides an overview of the model's fundamental aspects. Sec.~\ref{sec:HiggsDiracScoto} discusses the stability of the electroweak vacuum in this model. In Sec.~\ref{sec:CLFV}, we analyse the cLFV processes $\ell_\alpha \rightarrow \ell_\beta \gamma$ within this model and the prospects of detecting such processes in future experiments, as well as discuss how these processes can constrain the model. In Sect.~\ref{sec:DM}, we analyse the DM sector in the three limiting cases: doublet scalar DM, singlet scalar DM and fermionic DM. We find that all three DM cases are compatible with all the experimental constraints. We further show that the model allows for the possibility of novel low-mass DM, a feature not shared by its canonical Majorana counterpart\footnote{See \cite{Hirsch:2013ola} for an extension of the simplest Majorana scotogenic model which allows a lower mass DM.}. We conclude with final remarks in Sec.~\ref{sec:conclusions}.

\section{The Minimal Dirac Scotogenic Model}
\label{sec:model}
The Dirac Scotogenic model is a generic framework to obtain stable DM while simultaneously generating naturally small Dirac neutrino masses at the loop level \cite{Bonilla:2018ynb, Guo:2020qin,Hazarika:2022tlc,CentellesChulia:2022vpz}. This is typically accomplished by adding new symmetries to the SM, which then ensure the Dirac nature of neutrinos and/or the stability of the DM. An elegant way of doing so is by using the $B-L$ symmetry of SM and breaking it to an appropriate discrete subgroup which then simultaneously ensures the Diracness of neutrinos as well as the stability of DM as shown in Ref.~\cite{Bonilla:2018ynb}. In this work, we explore the phenomenological implications of a variant of the model presented in Ref.~\cite{Bonilla:2018ynb} with explicit breaking of global $B-L$ symmetry to its $\mathcal{Z}_6$ subgroup.

In the minimal simplified scenario, the SM gauge symmetries \SM are extended with a global $B-L$ symmetry which is then explicitly broken by soft terms to its $\mathcal{Z}_6$ subgroup. 
The particle content of our model and their corresponding charge assignments are given in Tab.~\ref{tab:model} 

\begin{table}[th]
\begin{center}
\begin{tabular}{| c || c | c  || c  c |}
  \hline 
& \hspace{0.1cm}  Fields  \hspace{0.1cm}          &  \hspace{0.1cm}  $SU(2)_L \otimes U(1)_Y$  \hspace{0.1cm} & \hspace{0.2cm}{\color{red}$U(1)_{B-L}$} \hspace{0.3cm} $\rightarrow$    & \hspace{0.15cm} {\color{blue}$\mathcal{Z}_{6}$}  \hspace{0.15cm}              \\
\hline \hline
\multirow{4}{*}{ \begin{turn}{90} \hspace{-0.15cm} Fermions \end{turn} } &
 $L_i$        	  &    ($\mathbf{2}, {-1/2}$)    & \hspace{0.1cm} {\color{red}$-1$} \hspace{1.3cm} $\rightarrow$ 	  &	 {\color{blue}$\omega^4$}                \\	
&   $\nu_{R_i}$       &   ($\mathbf{1}, {0}$)    & {\color{red}$(-4, -4, 5)$}  \hspace{0.2cm} $\rightarrow$   &  	 {\color{blue}$(\omega^4, \omega^4, \omega^4)$} \\
&   $N_{L_j}$    	  &   ($\mathbf{1}, {0}$)    & {\color{red}$-1/2$} \hspace{1.2cm} $\rightarrow$   &    {\color{blue} $\omega^5$}     \\
&  $N_{R_j}$     	  &  ($\mathbf{1}, {0}$) 	 & {\color{red}$-1/2$}  \hspace{1.2cm} $\rightarrow$    &  {\color{blue}$\omega^5$}    \\
\hline \hline
\multirow{4}{*}{ \begin{turn}{90} \hspace{0.2cm} Scalars \end{turn} } &
 $H$  		 &  ($\mathbf{2}, {1/2}$)            & \hspace{0.2cm} {\color{red}$0$} \hspace{1.5cm} $\rightarrow$     & {\color{blue} $1$}   \\
& $\eta$          	 &  ($\mathbf{2}, {1/2}$)    & \hspace{0.1cm} {\color{red}$1/2$}   \hspace{1.2cm} $\rightarrow$     &  {\color{blue}$\omega$}     \\
& $\xi$             &  $(\mathbf{1}, {0})$       & \hspace{0.1cm} {\color{red}$7/2$}  \hspace{1.2cm} $\rightarrow$     &	{\color{blue}$\omega$} \\	
    \hline
  \end{tabular}
\end{center}
\caption{Charge assignment for the lepton and scalar sector of our model. Here $\mathcal{Z}_6$ is the residual symmetry coming from $U(1)_{B-L}$ breaking, see \cite{Bonilla:2018ynb} for a detailed discussion. Note that all SM particles transform as even powers of $\omega$, with $\omega$ being the sixth root of unity, i.e. $\omega^6=1$. 
The particles of the dark sector, i.e. $N_L$, $N_R$, $\eta$ and $\xi$, all transform as odd powers of $\omega$   and the lightest of this set is the completely stable DM.}
 \label{tab:model}
\end{table}

The particle content of the SM gets extended by three light right handed neutrinos $\nu_{R_i}$; $i = 1,2,3$, two generations of Dirac heavy neutral fermions $N_{L_j}$ and $N_{R_j}$; $j = 1,2$, an inert doublet scalar $\eta$ and an inert singlet scalar $\xi$. This particle content leads to the mass generation of light Dirac neutrinos at the one-loop level, as illustrated in Fig. \ref{fig:numass}. All internal particles in the loop transform as odd powers of $\omega$ under the residual $\mathcal{Z}_6$ symmetry, while the SM particles transform as even powers of $\omega$. Consequently, any possible effective operator leading to the decay of a particle of the dark sector necessarily implies another dark sector particle. Therefore, the lightest of them is completely stable and is the DM candidate. Before addressing the neutrino mass generation as detailed in section ~\ref{sec:numass}, we first discuss the scalar sector of the model.
\subsection{The Scalar Sector}
\label{sec:scalar}
In the scalar sector, the Higgs field $H$ is SM-like. Additionally, we add a $SU(2)_L$ doublet $\eta$ and a $SU(2)_L$ singlet $\xi$. The general form of the scalar potential allowed by the symmetries of the model is given by

\begin{eqnarray}
V & = & -\mu_{H}^{2} H^{\dagger} H +\mu_{\eta}^{2} \eta^{\dagger} \eta + \mu_{\xi}^{2} \xi^{*} \xi
+\frac{1}{2} \lambda_{1}(H^{\dagger} H)^{2}+\frac{1}{2} \lambda_{2}(\eta^{\dagger} \eta)^{2}+\frac{1}{2} \lambda_{3}(\xi^{*} \xi)^{2} + \lambda_{4}(H^{\dagger} H)(\eta^{\dagger} \eta) \nonumber \\
& + &  \lambda_{5}(H^{\dagger} \eta)(\eta^{\dagger} H)+ \lambda_6(H^{\dagger} H)(\xi^{*} \xi) + \lambda_{7}(\eta^{\dagger} \eta)(\xi^{*} \xi)+ (\kappa \, \eta^{\dagger} H \xi + h.c.)
\label{eq:pot}
\end{eqnarray}

The soft term $\kappa \, \eta^{\dagger} H \xi + h.c$ explicitly breaks the global $U(1)_{B-L} \to \mathcal{Z}_6$. Since the symmetry of the Lagrangian gets enhanced when $\kappa\to 0$, it is a technically natural parameter and can be taken to be small. As we can see in Sec.~\ref{sec:numass}, the neutrino mass is proportional to $\kappa$. As such, in the Dirac scotogenic model, the mass-dimension parameter $\kappa$ plays the role that $\lambda_5$ plays in the Majorana version.

The requirement to have a stable minimum for the potential at tree level implies the following conditions in the scalar potential parameters \cite{Kannike:2016fmd}

\begin{eqnarray}
& & \lambda_{1}, \lambda_{2}, \lambda_{3}  >  0; \hspace{0.5cm} \lambda_{4} > -\sqrt{\lambda_{1}\lambda_{2}}, \hspace{0.5cm}  \lambda_6 > -\sqrt{\lambda_{1}\lambda_{3}}, \hspace{0.5cm} \lambda_{7} > -\sqrt{\lambda_{2}\lambda_{3}}, \nonumber\\
& &  \sqrt{\frac{\lambda_{1}}{2}}\lambda_{7} + \sqrt{\frac{\lambda_{2}}{2}}\lambda_6 + 
\sqrt{\frac{\lambda_{3}}{2}}\lambda_{4}  +\sqrt{\frac{\lambda_{1}\lambda_{2}\lambda_{3}}{8}} > -\sqrt{(\lambda_{4} +\sqrt{\lambda_{1}\lambda_{2}})(\lambda_6+\sqrt{\lambda_{1}\lambda_{3}})(\lambda_{7}+\sqrt{\lambda_{2}\lambda_{3}})} \hspace{0.3cm} .
\label{eq:vacuumstability}
  \end{eqnarray}

While the tree level perturbativity\footnote{See \cite{Krauss:2017xpj} for a more detailed analysis on loop-corrected perturbativity conditions.} of the dimensionless couplings implies

\begin{equation}
   |\lambda_j| \leq \sqrt{4 \pi}
    \label{eq:perturbativity1}
\end{equation}

with $j\in \{1 \cdots 7\}$. Fleshing out the $SU(2)_L$ components of the scalars, we can write
\begin{eqnarray}
& & \begin{aligned}
  H = \begin{pmatrix}
H^+\\
H^0
\end{pmatrix}, & \hspace{1cm}& \eta = \begin{pmatrix}
\eta^+\\
\eta^0
\end{pmatrix}
\end{aligned}\\
   & & H^0=\frac{1}{\sqrt{2}}(v+h+iA), \hspace{0.5cm} \eta^0=\frac{1}{\sqrt{2}}(\eta_{R}+i\eta_{I}), \hspace{0.5cm} \xi=\frac{1}{\sqrt{2}}(\xi_{R}+i\xi_{I}) \hspace{0.2cm}
\end{eqnarray}
Note that, as mentioned before, $H$ gets a vacuum expectation value (vev) while $\eta$ and $\xi$ are inert.

We can now compute the tree-level masses of the physical scalar states after symmetry-breaking. Note that the mixing between $H$ and $\eta$ or $\xi$ is forbidden by the $\mathcal{Z}_6$ symmetry.
\begin{equation}
    m_{h}^{2} = \lambda_{1}v^2,
\end{equation}
\begin{equation}
    m_{\eta^{\pm}}^{2} = \mu_{\eta}^{2} + \frac{\lambda_{4}}{2}v^2,
    \label{eq:metaplus}
\end{equation}

Owing to their $\mathcal{Z}_6$ symmetry charges, the scalar fields $\eta$ and $\xi$ are complex fields. The residual symmetry $\mathcal{Z}_6$ forces the neutral imaginary and real components of each dark sector scalar to be degenerate i.e. there will be no mass splitting between real and imaginary components of neutral dark scalar particles. However, the real part of $\xi$ mix with the real part of $\eta^0$ and similarly, the imaginary part of $\xi$ mix with the imaginary part of $\eta^0$ with the same mixing matrix.

\begin{equation}
    m_{(\xi_{R},\eta_{R})}^{2}=m_{(\xi_{I},\eta_{I})}^{2}=\begin{pmatrix}
    \mu_{\xi}^{2} + \lambda_6 \frac{v^2}{2} & \kappa \frac{v}{\sqrt{2}}\\
    \kappa \frac{v}{\sqrt{2}} & \mu_{\eta}^{2} + (\lambda_{4}+\lambda_{5}) \frac{v^2}{2}
    \end{pmatrix}
    \label{eq:metamix}
\end{equation}
We can analytically compute the singlet-doublet mixing angle $\theta$
\begin{equation}
    \tan2\theta=\frac{\sqrt{2}\kappa \, v}{ (\mu_{\xi}^{2}-\mu_{\eta}^{2})+(\lambda_6-\lambda_{4}-\lambda_{5})\frac{v^2}{2}} \ll 1
    \label{eq:mix}
\end{equation}

Since $\kappa$ is naturally small, the mixing between dark sector singlet and doublet is also extremely small. 
The mass eigenstates\footnote{To simplify notation, we denote the mass eigenstates as $\eta^0$ and $\xi$.} for the real/imaginary part of the neutral scalars are given by
\begin{eqnarray}
    m^2_{1R}=m^2_{1I}= \left(\mu_{\xi}^{2} + \lambda_6 \frac{v^2}{2}\right) \cos^2{\theta} + \left(\mu_{\eta}^{2} + (\lambda_{4}+\lambda_{5}) \frac{v^2}{2}\right)\sin^2{\theta} - 2\kappa v \sin{\theta} \cos{\theta}\equiv m^2_{\xi} \label{eq:scalarmass1}\\
m^2_{2R}=m^2_{2I}= \left(\mu_{\xi}^{2} + \lambda_6 \frac{v^2}{2}\right) \sin^2{\theta} + \left(\mu_{\eta}^{2} + (\lambda_{4}+\lambda_{5}) \frac{v^2}{2}\right)\cos^2{\theta} + 2\kappa v \sin{\theta} \cos{\theta}\equiv m^2_{\eta^0} \label{eq:scalarmass2}
\end{eqnarray}

In the small mixing angle limit, the first state is mainly a singlet while the second one is mainly a doublet. The lighter neutral mass eigenstate is stable and can be a good candidate for DM as discussed in Sec.~\ref{sec:DM}, if their mass is lighter than that of the heavy neutral fermion, $m_N$.

\subsection{Neutrino Mass Generation}
\label{sec:numass}

The $SU(3)_{c} \otimes SU(2)_{L} \otimes U(1)_{Y} \otimes U(1)_{B-L}$ invariant Yukawa Lagrangian relevant for the charged lepton and neutrino mass generation is given as

\begin{equation}
    -\mathcal{L}_{Y}\supset Y^\ell_{i j} \Bar{L}_{i} H e_{R_{j}} + Y_{i k} \Bar{L}_{i} \tilde{\eta} N_{R_{k}}+Y_{kl}^{\prime} \bar{N}_{L_{k}} \nu_{R_{l}} \xi+{M_N}_{k l} \bar{N}_{L_{k}} N_{R_{l}}+h . c.
    \label{eq:Yukawas}
\end{equation}

where $\tilde{\eta} = i \tau_{2} \eta^{*}$, $\tau_{2}$ is the $2^{nd}$ Pauli matrix and the indices $i, j \in \{1,2,3\}$, $k,l \in \{1, 2\}$. \\
$Y^\ell$, $Y$ and $Y^{\prime}$ are dimensionless Yukawa couplings of size $3\times 3$, $3\times 2$ and $2 \times 2$, respectively, while $M$ is a $2\times 2$ Dirac mass term for the neutral fermions $N_L$ and $N_R$. We can choose to work, without loss of generality, in the basis where both $Y^\ell$ and $M_N$ are diagonal and $Y'$ is hermitian. The tree level perturbativity implies

\begin{equation}
 Tr(Y^{\dagger} Y ) < 4\pi, \hspace{0.2cm} Tr(Y^{\prime \dagger} Y^{\prime} ) < 4\pi
 \label{eq:perturbativity2}
\end{equation}

\begin{figure}[th]
\includegraphics[height=4.0cm]{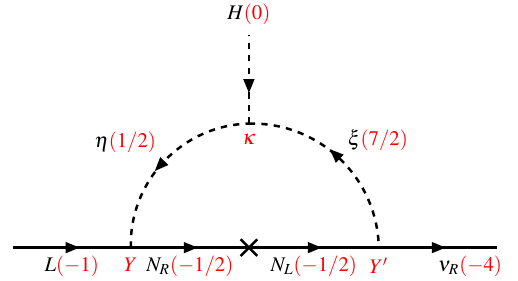}
\includegraphics[width=2cm]{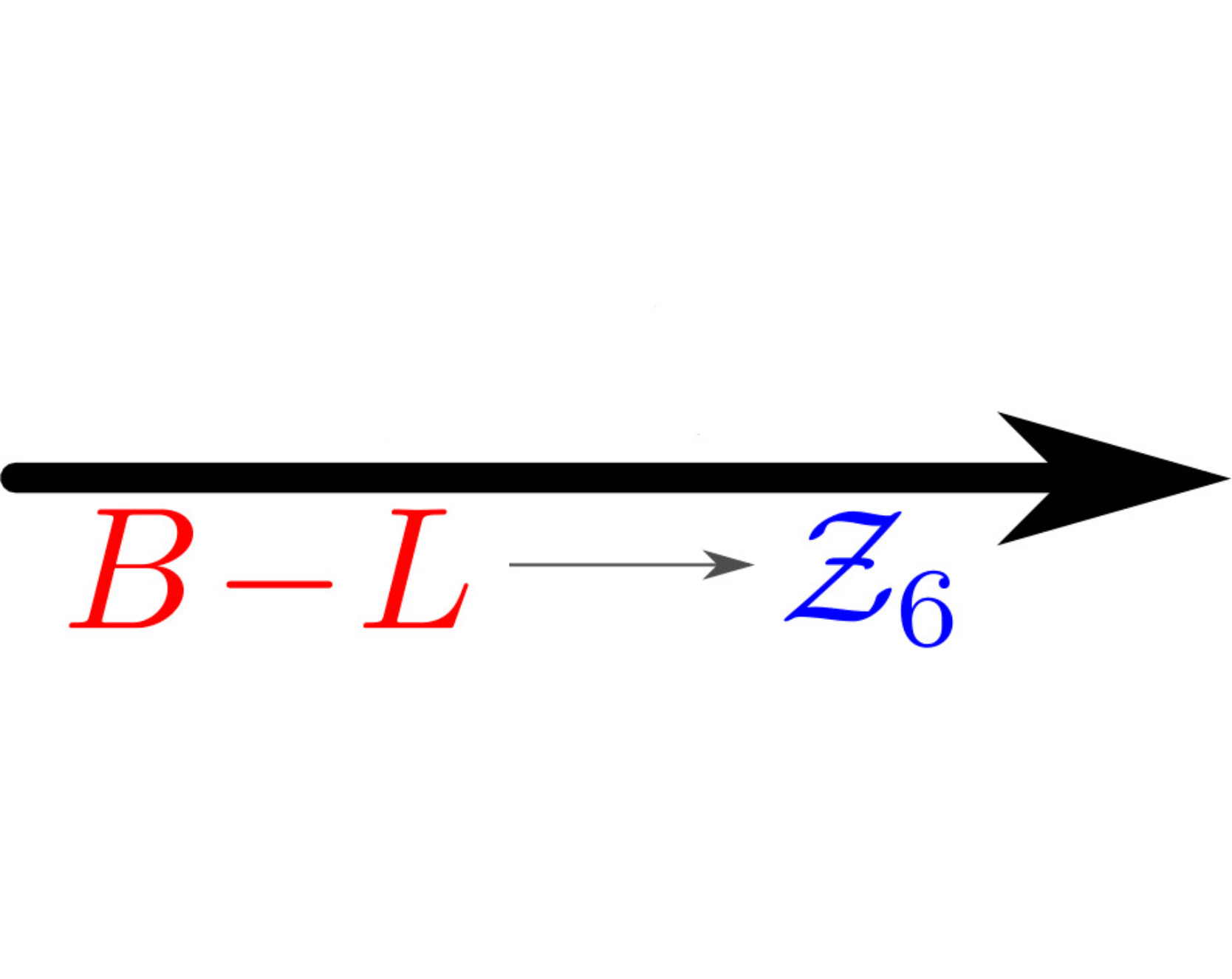}
\includegraphics[height=4.0cm]{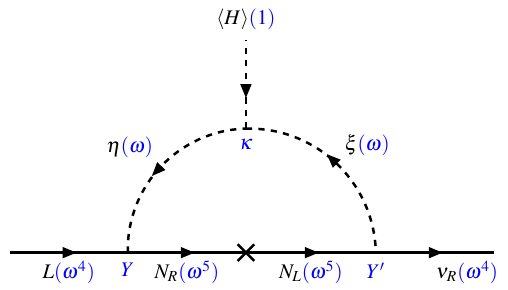}
        \caption{\begin{footnotesize}Leading neutrino mass generation diagram. The tree-level mass term between left and right-handed neutrinos is forbidden by the chiral $U(1)_{B-L}$ symmetry and only appears at the one-loop level after the $U(1)_{B-L} \to \mathcal{Z}_6$ symmetry breaking. The unbroken residual $\mathcal{Z}_6$ symmetry simultaneously ensures that neutrinos remain Dirac in nature and the lightest dark sector particle is completely stable. \end{footnotesize}}
        \label{fig:numass}
\end{figure}

The Lagrangian in Eq.~\ref{eq:Yukawas} generates a one-loop neutrino mass, as seen in Fig.~\ref{fig:numass}. We can calculate neutrino masses from the loop as

 \begin{equation}
    (M_\nu)_{ij} = \frac{1}{16\pi^2 \sqrt{2}} \sum\limits_{k=1}^2 Y_{ik} Y'_{kj} \frac{\kappa v}{\mu^2_\xi-\mu^2_\eta} M_{Nk} \left(\frac{m_{\xi}^2}{m_{\xi}^2-M_{Nk}^2}\log\frac{m_{\xi}^2}{M_{Nk}^2}-\frac{m_{\eta^0}^2}{m_{\eta^0}^2-M_{Nk}^2}\log\frac{m_{\eta^0}^2}{M_{Nk}^2}\right)
\end{equation}

$Y$, $Y'$ and $\kappa$ are the couplings described in Eq.~\ref{eq:Yukawas} and Fig.~\ref{fig:numass}, $\mu^2_\xi$ and  $\mu^2_{\eta}$ are the mass square parameters defined in Eq.~\ref{eq:pot}, $M_{Nk}$ are the heavy Dirac fermion mass, $m_\xi$ and $m_{\eta^0}$ are the neutral scalar mass eigenvalues of Eqs.~\ref{eq:scalarmass1} and~\ref{eq:scalarmass2}, and $v$ is the SM vev.
Note that this setup leads to only two massive neutrinos since $Y'$ is a rank-2 matrix. We can rewrite the neutrino mass matrix as
\begin{equation}
\label{eq:numass}
    M_{\nu} = Y M Y^{\prime}
\end{equation}

with the auxiliary diagonal matrix $M$ defined as
\begin{equation}
    M_{kl} = \delta_{kl} \frac{1}{16\pi^2 \sqrt{2}} \frac{\kappa v}{\mu^2_\xi-\mu^2_\eta} M_{N_k}\left(\frac{m_{\xi}^2}{m_{\xi}^2-M_{N_k}^2}\log\frac{m_{\xi}^2}{M_{N_k}^2}-\frac{m_{\eta^0}^2}{m_{\eta^0}^2-M_{N_k}^2}\log\frac{m_{\eta^0}^2}{M_{N_k}^2}\right)
\end{equation}

We now rotate the neutrino fields into their mass basis by $\nu_L \to U_L \nu_L$ and $\nu_R \to U_R \nu_R$, where $U_L$ is the PMNS matrix relevant to oscillations and $U_R$ is a $2\times 2$ unphysical unitary matrix. Since our setup is flavor-blind, it is clear that we can reproduce the mixing parameters and masses measured in neutrino oscillations \cite{deSalas:2020pgw}. The Yukawa matrices must satisfy

\begin{align}
    M_\nu^d = U_L^\dagger Y M Y' U_R & \rightarrow Y = U_L M_\nu^d U_R^\dagger Y'^{-1} M^{-1} \label{eqn:CIP}
\end{align}

where $M_\nu^d$ $3\times 2$ composed by a diagonal $2\times 2$ block with the neutrino masses and an extra row of zeros i.e. 1st row for Normal Ordering (NO) and 3rd row for Inverted Ordering (IO). The matrices $Y'$ and $M$ are $2\times 2$ and of rank $2$, so their inverse always exists. In the analysis that follows we impose neutrino masses and mixing inside the $3\sigma$ range of the global fit \cite{deSalas:2020pgw} and scan over the free parameters $Y'$, as well as the heavy neutral lepton masses $m_N$.
For the sake of representation and simplicity, we assume IO of neutrino masses in what follows. This is a natural choice, as the lightest neutrino is massless and the other two have a mass in the same scale, $(m_1,\, m_2,\, m_3) \approx (0.0495,\, 0.0502, \,0 )$ eV. However, we have explicitly checked that all of our results and conclusions also hold for NO.

\section{Higgs Vacuum Stability in the Dirac Scotogenic Model}
\label{sec:HiggsDiracScoto}

Before exploring various phenomenological implications, we first examine the vacuum stability of the model. The structure of the EW vacuum of the SM has been thoroughly studied and is known to be metastable, although with an exceptionally long lifetime \cite{Elias-Miro:2011sqh,Bezrukov:2012sa,Degrassi:2012ry,Masina:2012tz,Buttazzo:2013uya}. This may be seen as a hint of the presence of new physics at a higher scale that ensures the exact stability of the true vacuum of the model.

We now analyse the conditions under which the vacuum of the Dirac Scotogenic model can be stabilized up to the Planck scale. In our analysis, we work in the $\overline{MS}$ (Modified Minimal Subtraction) scheme and take the Renormalization Group (RG) running of all parameters up to the two-loop level.

In Table~\ref{tab:SM_RGE_Parameter}, we list the $\overline{MS}$ input values of the relevant parameters at the top quark mass $m_{t}$ scale and subsequently compute the Renormalization Group Equations (RGE) of the couplings of our model. The beta functions for various gauge, quartic, and Yukawa couplings in the model are evaluated up to the two-loop level using SARAH-4.14.5 \cite{Staub:2015kfa}.

\begin{table}[th]
\begin{center}
\begin{tabular}{|    c   |    c    | c | c | c | c |}
  \hline 
  Parameter    &   $g_{1}$   &   $g_{2}$    &   $g_{3}$ & $Y_{t}$   & $\lambda_{1}$ \\
\hline
$\mu(m_{t})$     &  	 0.462607            &  0.647737     &  	 1.16541 &  0.93519 & 0.256          \\	
    \hline
  \end{tabular}
\end{center}
\caption{$\overline{MS}$ values of the main input parameters at the top quark mass scale, $m_{t}$ = 172.69 ± 0.3 GeV. \cite{Mandal:2020lhl, ParticleDataGroup:2018ovx}}
 \label{tab:SM_RGE_Parameter} 
\end{table}
\FloatBarrier

In our analysis, we observe a notable dependence of the running of the quartic Higgs self-coupling ($\lambda_{1}$) on the interaction couplings $\lambda_{4}$, $\lambda_{5}$, and $\lambda_{6}$ parameters of the scalar potential in Eq.~\ref{eq:pot}.
 \begin{figure}[th]
 \centering
        \includegraphics[height=8cm]{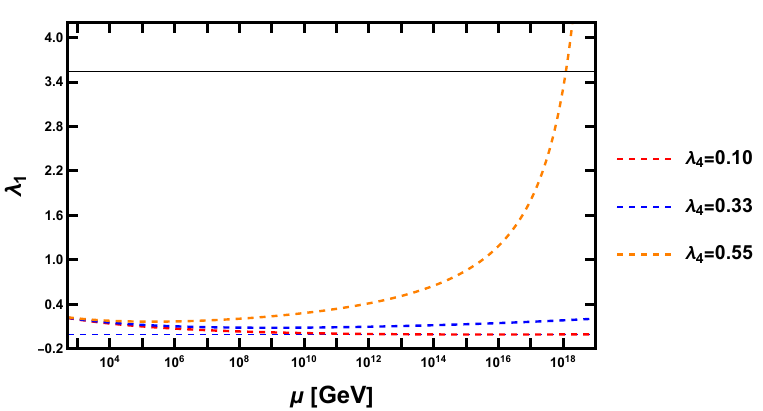}
        \caption{RG evolution of the quartic Higgs boson self-coupling $\lambda_{1}$ for different values of $\lambda_4$ while fixing $\lambda_5$ and $\lambda_6$. See text for details.}
        \label{fig:Eta_RGE_lhh}
\end{figure}
As an illustrative example, in Fig.~\ref{fig:Eta_RGE_lhh} we show the evolution of the Higgs self-coupling $\lambda_1$ for different initial values of the coupling $\lambda_4$, while fixing the initial values of $\lambda_5 = -1.0\times 10^{-2}$ and $\lambda_6 = -4.9\times 10^{-8}$. Setting the $\lambda_{4}$ coupling to 0.10 (red dashed line) does not provide sufficient correction to the $\lambda_1$ coupling, which still becomes negative at high energy scales. Increasing the value of $\lambda_4$ to 0.33 (blue dashed line) allows the $\lambda_1$ parameter to remain positive and within the perturbative regime up to the Planck energy scale. However, further increasing $\lambda_{4}$ to 0.55 (orange dashed line) results in the loss of perturbativity for $\lambda_1$ at some energy scale. A similar conclusion holds for the couplings $\lambda_5$ and $\lambda_6$. Of course while doing the above RG running, one also has to take the RG running of all other parameters into account. We have done this for the three possible type of DM
as we now discuss. 

The EW vacuum can remain completely stable for all three types of possible DM candidates: dark doublet scalar, dark singlet scalar and dark fermion, see Sec.~\ref{sec:DM} for details. To illustrate this, we have shown the full 2-loop RG running of all model parameters for three distinct benchmark scenarios corresponding to each type of DM candidate: The benchmark point B1 features a doublet scalar DM particle, while the points B2 and B3 include a singlet scalar and neutral fermion DM particles, respectively. In Tab.~\ref{tab:benchmark}, we show the values of all the relevant model parameters and physical observables at the EW scale ($\mu(m_t)$) for each benchmark point.

\begin{table}[th]
\begin{center}
\begin{tabular}{|    c   |    c    | c | c | c | c | c | c |}
  \hline 
  Parameter    &   B1   &   B2    &   B3 & Parameter & B1 & B2 & B3 \\
\hline
$\lambda_{1}$     &  	 0.255            &  0.255   &  0.254    & $m_{\eta^+}[\text{GeV}]$     &    355.83       &  225.87    &  164.60      \\
$\lambda_{2}$     &    $4.91 \times 10^{-4}$       &  $2.07 \times 10^{-5}$   &  $6.73 \times 10^{-6}$   &  $m_{N1}[\text{GeV}]$     &   341.93        &  210.92   &  164.06   \\
$\lambda_{3}$     &    $1.53 \times 10^{-6}$        & $1.4 \times 10^{-2}$   &  $1.92 \times 10^{-6}$    &  $m_{N2}[\text{GeV}]$     &   364.29        &  220.80   &  164.75   \\
$\lambda_{4}$     &     $0.31$      &  $0.3225$   &   $2.45 \times 10^{-6}$  & $m_{\nu_1}[\text{eV}]$     &     0.0496    	 &  0.0502   &   0.0492    \\
$\lambda_{5}$     &   $-0.33$        &  $-0.32$   &  $-2.74 \times 10^{-4}$   &  $m_{\nu_2}[\text{eV}]$     &     0.0504    	 &  0.0510   &    0.0500   \\
$\lambda_{6}$     &   $-4.9 \times 10^{-8}$        &  $-7.14 \times 10^{-3}$   &  $0.35$    & $m_{\nu_3}[\text{eV}]$     &     0    	 &  0   &   0   \\
$\lambda_{7}$     &   $1.42 \times 10^{-3}$   &  $-4.13 \times 10^{-7}$   &  $1.36 \times 10^{-7}$    & $\theta_{12}[^{\circ}]$     &     36.47    	 & 31.85    &   34.89   \\
$\kappa [\text{GeV}]$          &    $0.50$       &      $1.46 \times 10^{-2}$  &  $-1.99$ &   $\theta_{13}[^{\circ}]$     &     8.29    	 &  8.40   &   8.72  \\
$\theta$          &    $2.4 \times 10^{-3}$       &      $-2.12 \times 10^{-3}$    &  $-1.66 \times 10^{-3}$ & $\theta_{23}[^{\circ}]$     &     46.58    	 &  43.12   &  44.77   \\
$\mu^2_{\eta}[\text{GeV}^2]$     &      $1.17 \times 10^{5}$     &  $4.09 \times 10^{4}$   &  $2.69 \times 10^{4}$    & $\Omega h^2$         &	  0.1151          &  0.1220  &  0.1180   \\
$\mu^2_{\xi}[\text{GeV}^2]$     &       $1.53 \times 10^{5}$     & $4.00 \times 10^{4}$     &  $2.15 \times 10^{5}$    &  $\sigma^{SI} [cm^{2}]$             &	 $1.70 \times 10^{-47}$           &  $6.68 \times 10^{-48}$  &  0   \\
  $m_{h}[\text{GeV}]$     &     125.33    	 &  125.19   &   125.27  & BR($\mu \rightarrow e \gamma$)             &	 $1.51 \times 10^{-13}$           &  $1.13 \times 10^{-13}$  &  $2.77 \times 10^{-13}$  \\	
	$m_{\eta^0}[\text{GeV}]$    &    340.84       &  202.48 &  164.27  & BR($\tau \rightarrow e \gamma$)             &	 $2.08 \times 10^{-13}$           &  $4.32 \times 10^{-14}$  &  $1.49 \times 10^{-12}$ 	 \\	
 $m_{\xi}[\text{GeV}]$    &     391.34      &   199.34    &  474.68 & BR($\tau \rightarrow \mu \gamma$)             &	 $2.27 \times 10^{-14}$           &  $6.29 \times 10^{-15}$  &  $2.94 \times 10^{-12}$	 \\
    \hline
  \end{tabular}
\end{center} 
\caption{Relevant model  parameters for three representative benchmark points, B1 (doublet scalar DM), B2 (singlet scalar DM) and B3 (fermionic DM) at the EW scale used for the RG evolution of the couplings.}
 \label{tab:benchmark} 
\end{table}

\begin{figure}[h!]
 \centering
    \includegraphics[height=4.0cm]{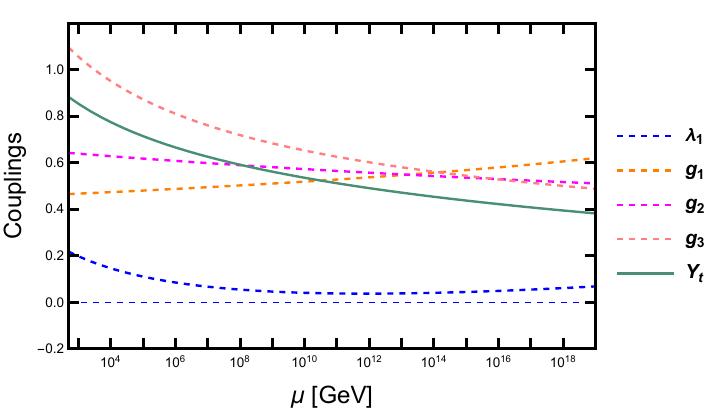}
    \includegraphics[height=4.0cm]{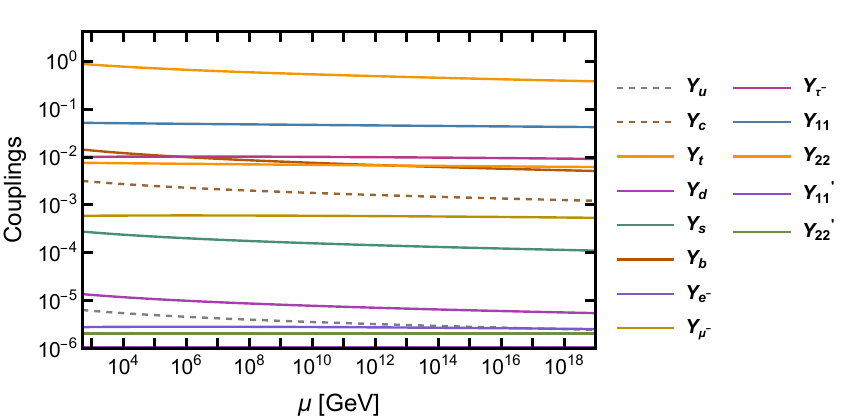}
    \includegraphics[height=4.0cm]{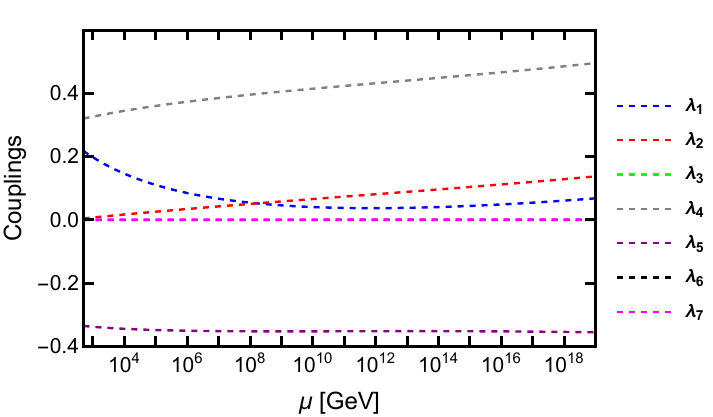}
        \caption{RG evolution of the relevant parameters of the benchmark point B1, featuring a doublet scalar DM particle.  \textbf{Top Left:} SM gauge couplings $g_{1}$, $g_{2}$, $g_{3}$, top quark Yukawa
coupling $Y_{t}$ and quartic Higgs boson self-coupling $\lambda_{1}$, \textbf{Top Right:} SM Yukawas and Neutrino Yukawas.  \textbf{Bottom:} Scalar potential quartic couplings $\lambda_i$}
        \label{fig:Eta_RGE}
\end{figure}

\begin{figure}[h!]
 \centering
        \includegraphics[height=4.0cm]{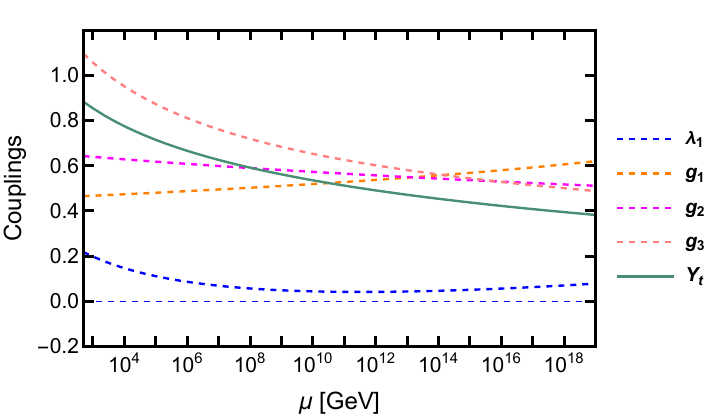}
          \includegraphics[height=4.0cm]{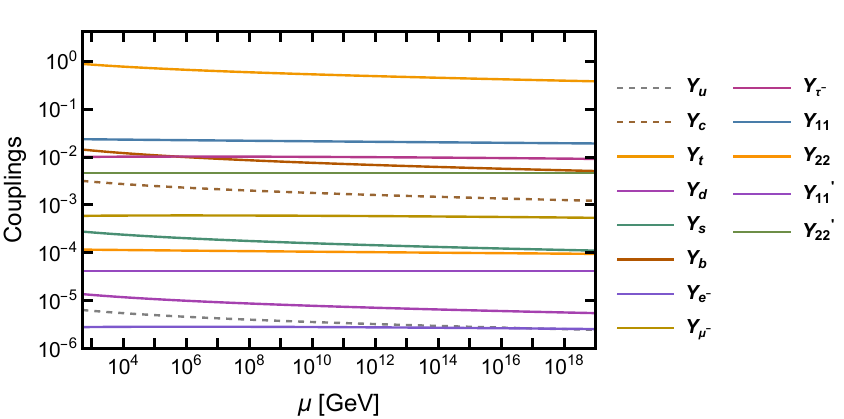}
           \includegraphics[height=4.0cm]{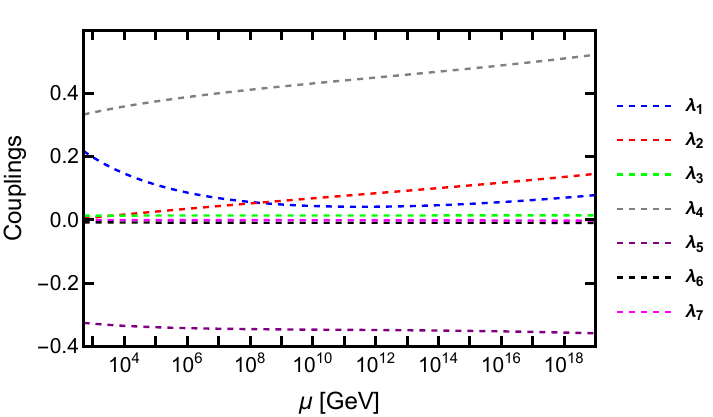}
        \caption{RG evolution of the relevant parameters of the benchmark point B2, featuring a singlet scalar DM particle. See Fig.~\ref{fig:Eta_RGE} for more details.}
        \label{fig:Xi_RGE}
\end{figure}

\begin{figure}[h!]
 \centering
       \includegraphics[height=4.0cm]{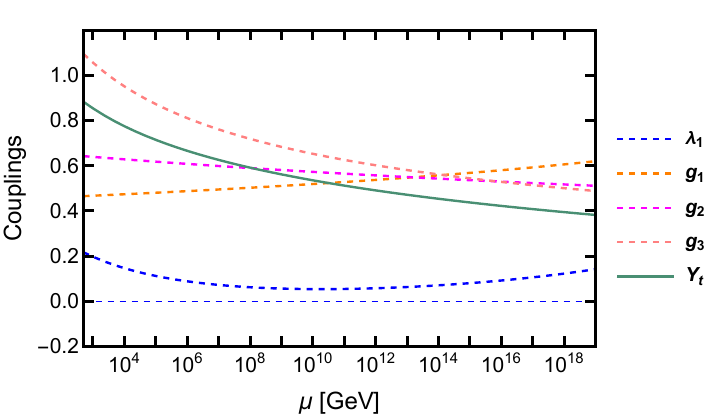}
          \includegraphics[height=4.0cm]{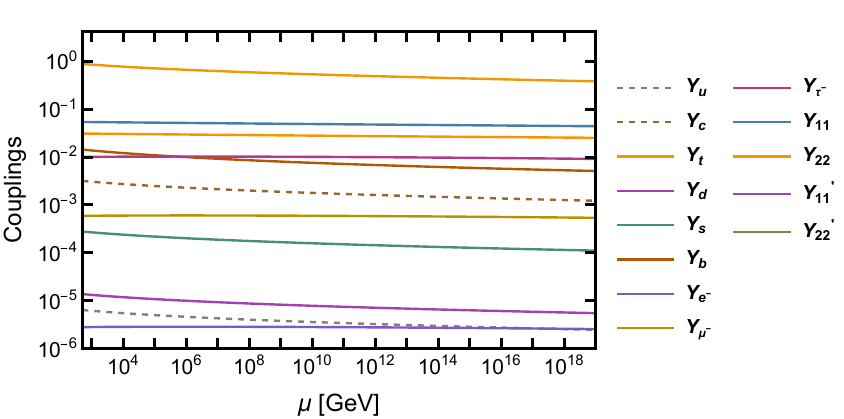}
          \includegraphics[height=4.0cm]{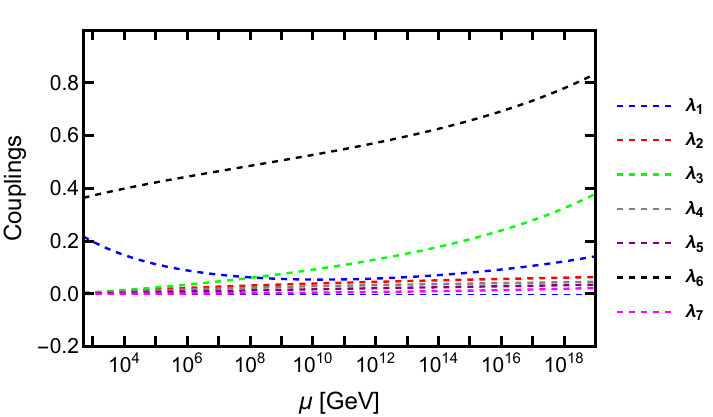}
        \caption{RG evolution of the relevant parameters of the benchmark point B3, featuring a fermion DM particle. See Fig.~\ref{fig:Eta_RGE} for more details.}
        \label{fig:N1_RGE}
\end{figure}

All three benchmark points are selected such that the perturbativity and stability conditions of \eqref{eq:vacuumstability}, \eqref{eq:perturbativity1} and \eqref{eq:perturbativity2} are satisfied, as well as the relevant phenomenological constraints from colliders, relic density and cLFV, see details in Secs.~\ref{sec:CLFV} and \ref{sec:DM}. In Figs.~\ref{fig:Eta_RGE}, \ref{fig:Xi_RGE} and \ref{fig:N1_RGE}, we show the RG evolution of various couplings. 

In the left panel of the upper row of each of the three plots, we show the RG evolution of the SM gauge couplings, the top quark Yukawa coupling and the quartic Higgs boson self-coupling. The gauge couplings become approximately same strength (but do not unify) at a scale of around $10^{14}$ GeV. The quartic Higgs boson coupling $\lambda_1$ remains positive and within the perturbative regime across all energy scales for the three cases, while the RG evolution of the top quark Yukawa coupling closely mirrors that of the SM.

In each of the top-right panels, we show the RG evolution of the
SM Yukawas and neutrino Yukawa couplings\footnote{For sake of not cluttering the plots too much, we are only showing the running of $Y_{11}, Y_{22}, Y^{\prime}_{11}$ and $Y^{\prime}_{22}$ components of neutrino Yukawas .}, which are not significantly affected by the running of the scale. Finally, in the lower row of each three plots, we show the RG evolution of the quartic couplings of the scalar potential $\lambda_i$. As seen from all three figures, the Higgs vacuum is stable as all self-couplings $(\lambda_{1}, \lambda_{2}$ and $\lambda_{3})$ are positive and within their perturbative regime. All other additional BSM couplings can go negative, but maintain their value as defined by Eq.~\ref{eq:vacuumstability} at all energy scales $\mu$. We observe that these couplings increase with the energy scale, while remaining within the perturbative regime. Hence, the Higgs vacuum remains stable as the quartic couplings of the scalar potential satisfy their values by Eq.~\ref{eq:vacuumstability} for all energy values as shown in Fig.~\ref{fig:Eta_RGE},\ref{fig:Xi_RGE} and \ref{fig:N1_RGE}.

\FloatBarrier

\FloatBarrier

While a detailed parameter scan of the Higgs vacuum stability and running of the parameters is outside the scope of this work, we have shown by taking benchmark scenarios demonstrating that the vacuum becomes absolutely stable in the context of the Dirac Scotogenic model, while satisfying the relevant phenomenological constraints at the same time. 

\section{Charged Lepton Flavor Violation}
\label{sec:CLFV}
The SM with massless neutrinos predicts lepton flavor conservation. However, once neutrinos become massive lepton flavor can be violated in neutrino oscillations, where neutrinos change between different flavor states as they propagate through space. Indeed, many neutrino mass models, in particular the low scale constructions \cite{Akhmedov:1995ip,Akhmedov:1995vm,Malinsky:2005bi,Mohapatra:1986bd,Gonzalez-Garcia:1988okv}, generically predict sizeable rates of cLFV processes. For that reason, intense experimental \cite{Jodidio:1986mz,MEG:2016leq, COMET:2018auw,Belle:2021ysv,Moritsu:2022lem,Xing:2022rob, MEGII:2023ltw,Perrevoort:2024qtc} efforts have been and shall be dedicated to investigate various flavor-violating decays of muons and tau leptons. While the vast majority of such constructions lead to Majorana neutrinos, it is well-known that the existence of such processes does not have any implication on the Dirac/Majorana nature of neutrinos, since they can be present even with exactly massless neutrinos \cite{Bernabeu:1987gr}.

The strongest constraints to cLFV come from the family of non-standard lepton decays $\ell_{\alpha} \rightarrow \ell_{\beta} \gamma$, where $\alpha$ and $\beta$ represent different flavors of charged leptons. The current upper limits on their branching ratios are set as $BR(\mu \rightarrow e \gamma) < 3.1 \times 10^{-13} $ \cite{MEGII:2023ltw}, $BR(\tau \rightarrow e \gamma) < 3.3 \times 10^{-8}$ \cite{BaBar:2009hkt}, and $BR(\tau \rightarrow \mu \gamma) < 4.2 \times 10^{-8}$ \cite{Belle:2021ysv}, respectively at 90$\%$ confidence level (C.L.). In the near future, these limits are expected to be improved, with projected limits of $BR(\mu \rightarrow e \gamma) < 6 \times 10^{-14}$  \cite{MEGII:2021fah}, $BR(\tau \rightarrow e \gamma) < 9 \times 10^{-9}$ \cite{Belle-II:2022cgf}, and $BR(\tau \rightarrow \mu \gamma) < 6.9 \times 10^{-9}$ \cite{Belle-II:2022cgf}, respectively.

\begin{figure}[ht]
\includegraphics[width=8cm]{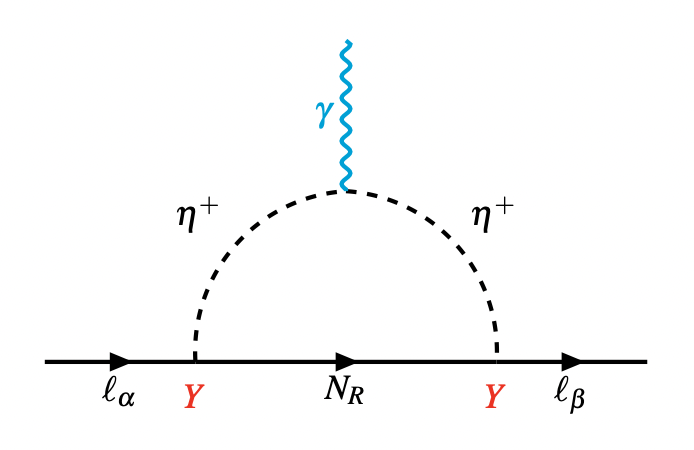}
        \caption{One-loop Feynman diagram for the process $\ell_{\alpha} \rightarrow \ell_{\beta} \gamma$}
        \label{fig:LFV}
\end{figure}

In the model discussed here, the leading contribution to $\ell_{\alpha} \rightarrow \ell_{\beta} \gamma$ comes at the one-loop level through the mediation of the charged scalar $\eta^+$ as shown in the diagram of Fig.~\ref{fig:LFV}.
Note that the neutrino mass is suppressed by the symmetry-protected $\kappa$ parameter, see the discussion in Sec.~\ref{sec:numass}. However, $\kappa$ does not enter the cLFV processes; therefore, our model can lead to sizeable cLFV even in the (unphysical) limit $\kappa \to 0$ implying $m_\nu \to 0$. 
A detailed calculation can be found in Appendix \ref{sec:appendix2}. The branching ratio of this process is given by 

\begin{equation}   
\operatorname{Br}(\ell_{\alpha} \rightarrow \ell_{\beta} \gamma)=\operatorname{Br}\left(\ell_{\alpha} \rightarrow \ell_{\beta} \nu_{\alpha} \overline{\nu_{\beta}}\right) \times \frac{3 \alpha_{e m}}{16 \pi G_{F}^{2}}\left|\sum_{i} \frac{Y_{\beta i}Y^{*}_{\alpha i}}{m_{\eta^{-}}^{2}} j\left(\frac{M_{N_{i}}^{2}}{m_{\eta^{-}}^{2}}\right)\right|^{2}
\label{eqn:cLFVeqn}
\end{equation}
and we take the numerical  values of $\operatorname{Br}\left(\ell_{\alpha} \rightarrow \ell_{\beta} \nu_{\alpha} \overline{\nu_{\beta}}\right)$, $\alpha_{em}$ and $G_F$  from the PDG \cite{ParticleDataGroup:2022pth}. The loop function $j(x)$ is given by:
\begin{equation}   
j(x)=\frac{1-6x+3x^{2}+2x^{3}-6x^{2}\log(x)}{12(1-x)^{4}}
\end{equation}

Note that among the three possible cLFV decays, the $\mu \to e \gamma$ has the most stringent experimental bound \cite{MEGII:2023ltw} and it constraints most of the parameter space compared to tau decays. 

Our setup is flavor blind and therefore we do not expect strong hierarchies in the Yukawa matrix $Y$. Moreover, since $Y^\prime$ does not enter the cLFV rates, mixing angles and neutrino masses can always be fitted for any arbitrary $Y$, see Eq.~\ref{eqn:CIP}. On the other hand, note that the masses of both the neutral fermion and the charged scalar masses, relevant for the DM phenomenology as will be discussed in Sec.~\ref{sec:DM}, appear in Eq.~\ref{eqn:cLFVeqn}. We now explore this interplay between cLFV process and DM for the three possible DM cases. The most important input parameters range are taken as given in the Tab.~\ref{tab:cLFVparameter}, while the other model parameters range are given in Tab.~\ref{tab:parameterrange}. In addition we have also imposed all experimental constraints listed in Sec.~\ref{sec:DM}. 
By using the analytical expression of eqn.~\ref{eqn:cLFVeqn} and input parameters range given in Tab.~\ref{tab:cLFVparameter} and ~\ref{tab:parameterrange}, we can analyse the cLFV for three different DM candidates.

\begin{table}
\begin{center}
\begin{tabular}{|    c   |    c    | c | c |}
  \hline 
  Parameter    &   Range   &   Parameter    &   Range  \\
\hline
$m_{\eta^{\pm}}$  &  $[10,10^{4}]\text{ GeV}$       &  $Y_{ij}$     & $[10^{-11},\sqrt{4\pi}]$\\
$M_{N_1}$ & $[10,10^{4}] \text{ GeV}$ 	 & 
$M_{N_2}$ & $[10,10^{4}] \text{ GeV}$  \\	
    \hline
  \end{tabular}
\end{center}
\caption{Value range for the numerical parameter scan for cLFV. Other model parameters range are given in Tab.~\ref{tab:parameterrange}.}
 \label{tab:cLFVparameter} 
\end{table}

\subsection{cLFV for Doublet Scalar DM}
We start with the case where the neutral component ($\eta^0$) of the doublet $\eta$ is the lightest dark sector particle i.e. the dark matter. In this case the mass of $\eta^\pm$ is slightly larger but close to mass of the DM. The dark fermion masses $m_{N_i}$; $i = 1,2$ are freely varied in the range mentioned in Tab.~\ref{tab:cLFVparameter} always ensuring that their masses are larger than DM mass. 

\begin{figure}[h!]
\centering
        \includegraphics[height=8.5cm]{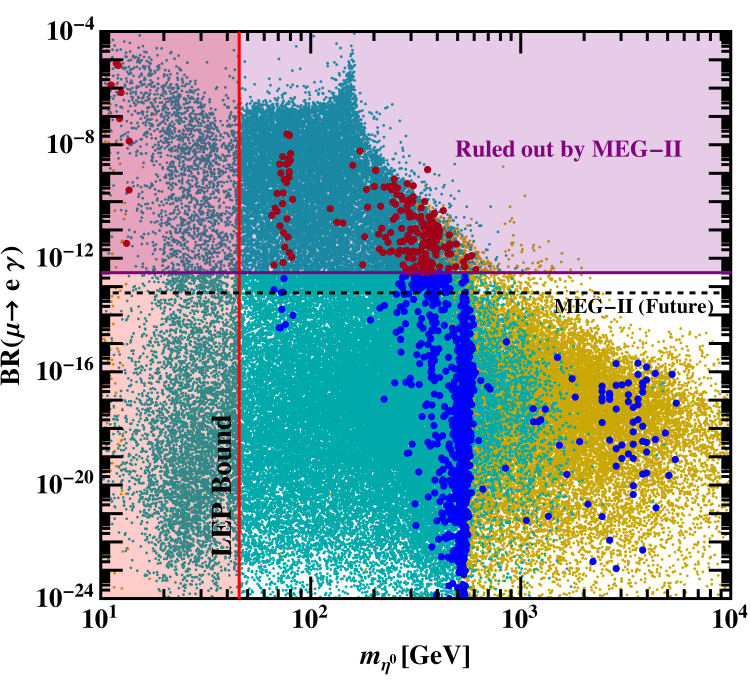}
            \caption{BR($\mu \rightarrow e \gamma$) vs doublet scalar DM mass. The yellow/cyan points represent over/under-abundant relic density \cite{Planck:2018vyg}, respectively. Red points satisfy the correct DM relic density but violate cLFV constraints. Blue points satisfy all constraints. The current experimental limits and possible future reaches are shown by purple shaded and black dashed lines respectively.  
        }
        \label{fig:mu-Eta_LFV}
\end{figure}
\begin{figure}[h!]
\centering
\includegraphics[height=7.5cm]{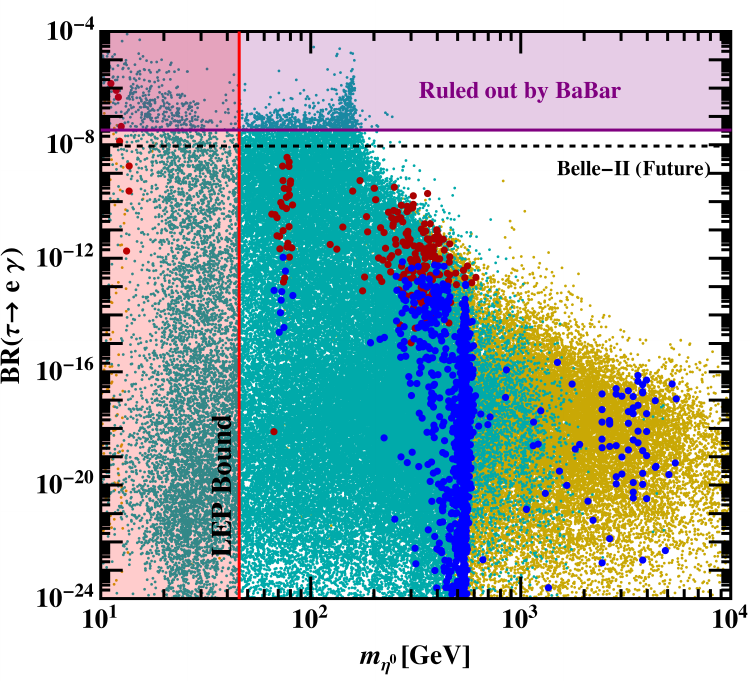}
      \includegraphics[height=7.5cm]{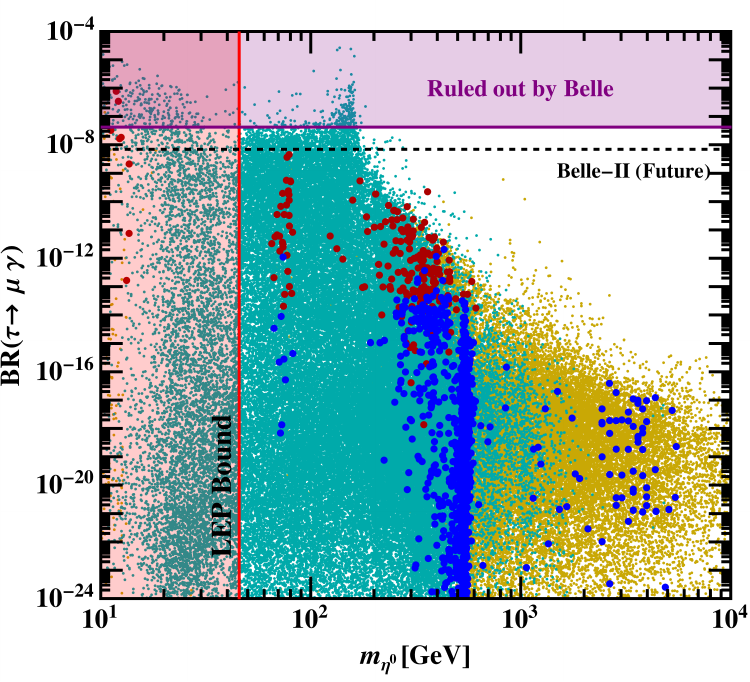}
  \caption{\textbf{Left panel:} BR($\tau \rightarrow e \gamma$) vs doublet scalar DM mass. \textbf{Right panel:} BR($\tau \rightarrow \mu \gamma$) vs doublet
        DM mass. The color code is the same as in Fig.~\ref{fig:mu-Eta_LFV}.}
        
        \label{fig:tau-Eta_LFV}
\end{figure}

Since, $ \mu \to e \gamma$ has most stringent experimental bounds, we start with analyzing it first. In Fig.~\ref{fig:mu-Eta_LFV}, we show the resulting $ \mu \to e \gamma$ decay rate as a function of the doublet scalar DM mass. Similarly, the cLFV decays of $tau$ lepton as a function of the doublet scalar DM mass are plotted in Fig. \ref{fig:tau-Eta_LFV}.
For comparison, current limits at 90$\%$ C.L. are also shown in the figure as a solid purple line \cite{MEGII:2023ltw, BaBar:2009hkt, Belle:2021ysv}, as well as the expected future sensitivity for all processes as dashed black line \cite{MEGII:2021fah, Belle-II:2022cgf}. 

As can be seen in Fig.~\ref{fig:mu-Eta_LFV}, we obtained cLFV within the current experimental bounds over a wide range of parameter space.
Note that all three processes feature similar behaviour, where their maximum achievable values for a given mass decrease as the DM mass increases. This feature occurs because, in this model, DM is always the lightest dark sector particle. Therefore,  as the mass of DM increases, the masses of the other particles within the dark sector also increases, resulting in a reduction in the cLFV rates, as described by Eq. \ref{eqn:cLFVeqn}. 
Since in our model the neutrino and DM are intertwined; therefore, the DM sector imposes severe constraints on cLFV.  In Fig.~\ref{fig:mu-Eta_LFV} and Fig.~\ref{fig:tau-Eta_LFV}, the yellow and cyan points represent regions of DM over-abundance and under-abundance, respectively, and are thus excluded. Red points satisfy the correct DM relic density but violate cLFV constraints. Only the blue points are allowed after imposing DM relic density as well the LEP constraints, Higgs boson mass, cLFV bounds and Higgs invisible branching ratio at 95$\%$ C.L.
etc. as discussed in Sec.~\ref{sec:DM}.
The blue points span three distinct mass regions: a low-mass region between half the Higgs mass and the $W$-boson mass, a medium-mass region between 200 and 550 GeV, and a high-mass region in the TeV range. These regions correspond to the blue points of Fig. \ref{fig:doubletDMhierarchichal} and Fig. \ref{fig:doubletDMcoannihilation} of Sec.~\ref{sec:doubletDM}.
Among the blue points, the high-mass region exhibits cLFV rates that surpasses the sensitivity of current and upcoming experiments.

However, the low and medium mass regions fall within the current experimental bounds and projected future limits for the $\mu \rightarrow e \gamma$ decay. In contrast, for the $\tau \to e \gamma$ and $\tau \to \mu \gamma$ decay, we do not obtain any mass region that falls within the current experimental and projected future limits as shown in the left and right panel of Fig.~\ref{fig:tau-Eta_LFV}. Thus, in this model, cLFV and DM constraints are complementary to one another. It is important to mention that our doublet scalar DM results are analogous to the result in \cite{Guo:2020qin} where the relation between neutrino Yukawa parameter and mass of DM is given for cLFV rates. However, we have also analysed the results for singlet scalar DM that we present in the next section.

\subsection{cLFV for Singlet Scalar DM}

\begin{figure}[h!]
\centering
        \includegraphics[height=8.5cm]{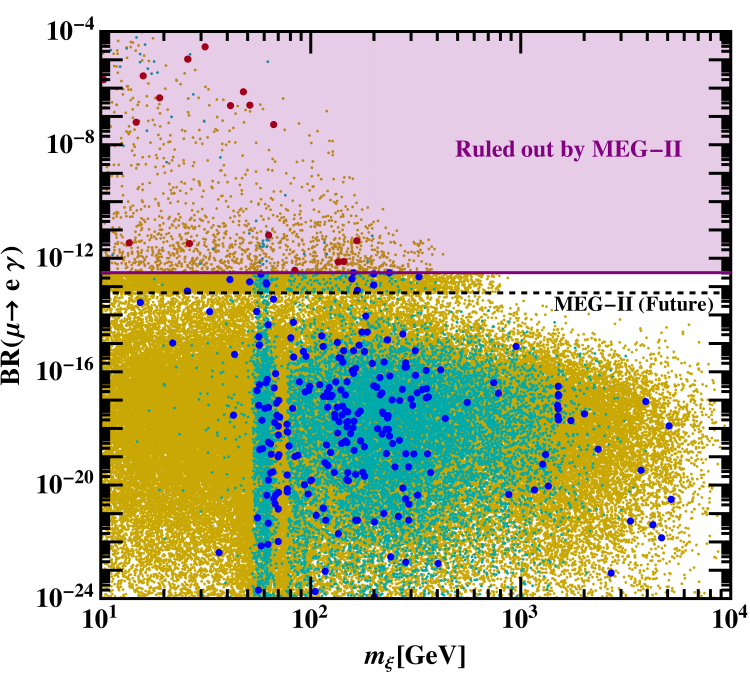}
        \caption{ BR($\mu \rightarrow e \gamma$) vs singlet scalar DM mass. The color code is the same as in Fig.~\ref{fig:mu-Eta_LFV}.}
        \label{fig:mu-Xi_LFV}
\end{figure}

\begin{figure}[h!]
\centering
    \includegraphics[height=7.5cm]{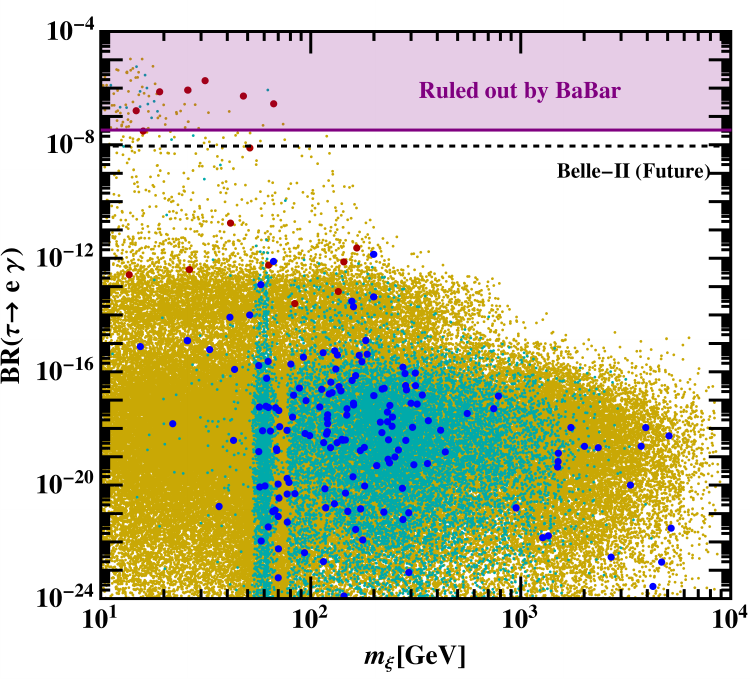}
      \includegraphics[height=7.5cm]{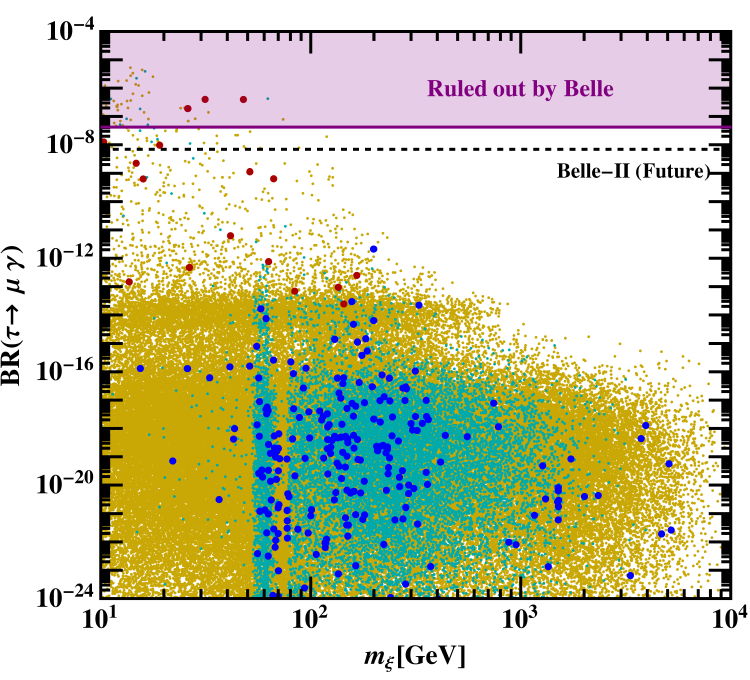}
        \caption{\textbf{Left panel:} BR($\tau \rightarrow e \gamma$) vs singlet
        DM mass. \textbf{Right panel:} BR($\tau \rightarrow \mu \gamma$) vs singlet
        DM mass. The color code is the same as in Fig.~\ref{fig:mu-Eta_LFV}.}
        \label{fig:tau-Xi_LFV}
\end{figure}

Now, we move on to the case, where the singlet scalar $(\xi)$ is the lightest dark sector particle. In Fig.~\ref{fig:mu-Xi_LFV} and ~\ref{fig:tau-Xi_LFV}, we show the resulting cLFV decay rates as a function of the singlet scalar DM mass. The color code is the same as in the previous section.
Compared to the doublet scalar DM case, the branching ratios of the cLFV decays tend to be lower, because $\eta^{\pm}$ and neutral fermion masses need to be higher than the singlet scalar mass as singlet is now the lightest dark sector particle, thus suppressing the cLFV rates. 

As discussed in the previous section, blue points span a wide range of parameter space. Again there are distinct patches where the DM relic density is satisfied, corresponding to co-annihilation dominated points at lower DM mass and hierarchical dark sector masses at higher DM mass regime as can be seen by relic abundance plots in Sec. \ref{sec:singletDM}.
The lower and medium mass region fall within the current experimental and projected future limits for the $\mu \to e \gamma$ decay. In contrast, for the $\tau \to e \gamma$ and $\tau \to \mu \gamma$ decay, we do not obtain any mass region that fall within the current experimental and projected future limits as shown in the left and right panel of Fig.~\ref{fig:tau-Xi_LFV}.

\FloatBarrier
\subsection{cLFV for Fermionic DM}

\begin{figure}[h!]
\centering
        \includegraphics[height=8.5cm]{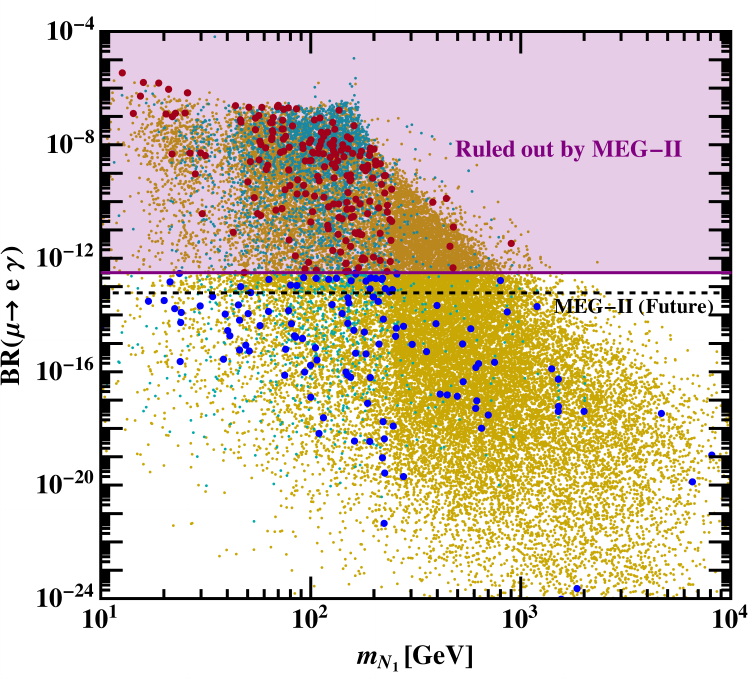}
        \caption{ BR($\mu \rightarrow e \gamma$) vs fermion DM mass. The color code is the same as in Fig.~\ref{fig:mu-Eta_LFV}.}
        \label{fig:mu-N1_LFV}
\end{figure}

\begin{figure}[h!]
\centering
    \includegraphics[height=7.5cm]{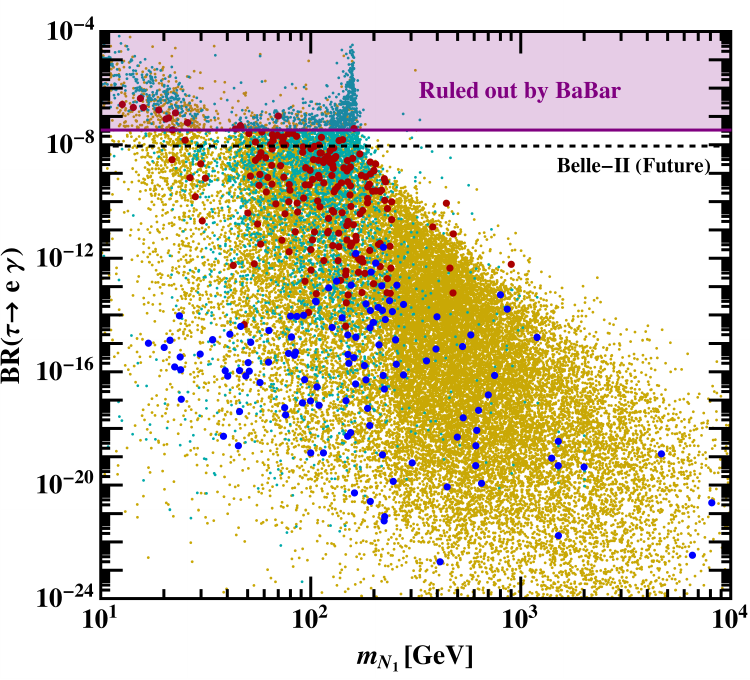}
      \includegraphics[height=7.5cm]{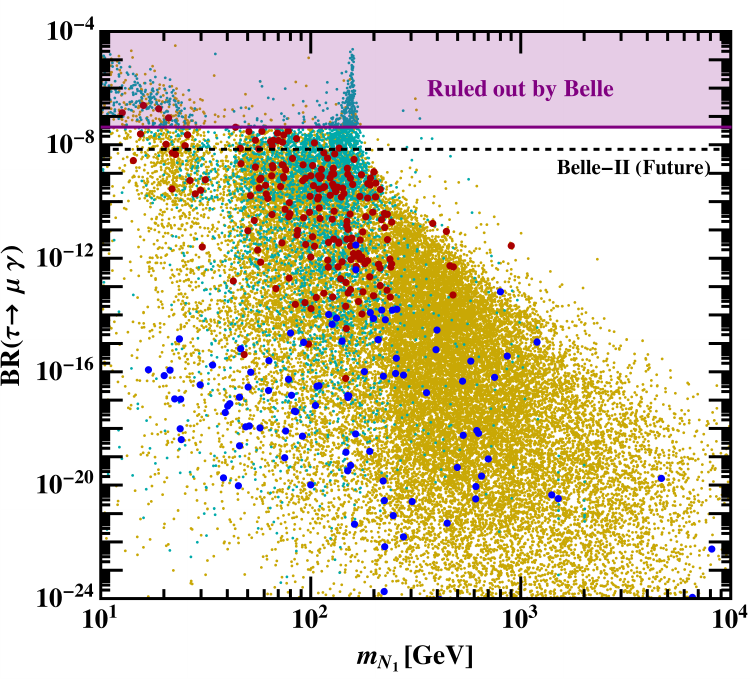}
        \caption{\textbf{Left panel:} BR($\tau \rightarrow e \gamma$) vs fermion
        DM mass. \textbf{Right panel:} BR($\tau \rightarrow \mu \gamma$) vs fermion
        DM mass. The color code is the same as in Fig.~\ref{fig:mu-Eta_LFV}.}
        \label{fig:tau-N1_LFV}
\end{figure}

Finally, we analyse the last case where the DM is the $N_1$ dark fermion. In Fig.~\ref{fig:mu-N1_LFV} and ~\ref{fig:tau-N1_LFV}, we show the resulting cLFV decay rates as a function of the fermionic DM mass. The color code is the same as in the previous section.

Compared to the above two cases, the cLFV rates tend to be higher in this case. This is because for fermionic DM, the relic density can only be satisfied through co-annihilation between DM and dark scalars, see Sec. \ref{sec:fermionDM}. This requires the singlet fermion and at least one of the  dark scalar masses to be nearly degenerate. In case when the DM is nearly degenerate with doublet scalar mass there is a double enhancement in cLFV rates from Eq.~\ref{eqn:cLFVeqn} leading to large cLFV rates.

In our analysis, we have identified data points within the wide range, which lie between the current experimental bounds and the anticipated future limits for the $\mu \rightarrow e \gamma$ decay, while at the same time respecting the DM relic density shown in Sec.~\ref{sec:fermionDM}. In contrast, for the $\tau \to e \gamma$ and $\tau \to \mu \gamma$ decay, we do not obtain any mass region that fall within the current experimental and projected future limits as shown in the left and right panel of Fig.~\ref{fig:tau-N1_LFV}. 
These results are analogous to the results in \cite{Guo:2020qin}, where the relation between neutrino Yukawa parameters and fermion DM mass is presented. 

In conclusion for all the three DM cases, there is an interesting interplay between cLFV and DM phenomenology. In particular when DM is the doublet scalar or the fermion, the cLFV rates are quite enhanced and are well within the reach of current and upcoming experiments. However, the constraints coming from DM sector also rule out a large parameter space with cLFV in experimental reach. The cLFV points within the reach of experiments and also allowed from DM constraints only occur in few distinct patches and thus cLFV measurements can be used to test our model. 

\section{Dark Matter Phenomenology}
\label{sec:DM}

We now focus on the analysis of the dark sector, with the lightest particle of the dark sector being completely stable due to the residual $\mathcal{Z}_6$ symmetry. This means that once such stable particle decouples from the thermal bath, its relic density `freezes out'. The relic density can then be computed and compared with Planck observations \cite{Planck:2018vyg}, $0.1126 \leq \Omega h^2 \leq 0.1246$ at the 3$\sigma$ confidence level \cite{Batra:2023bqj}. The production-annihilation diagrams used for computation of relic density are shown in the Appendix \ref{sec:appendix}. The DM in our model could also be detected by direct detection experiments such as XENONnT \cite{XENON:2023cxc}, LZ\footnote{We have also included the 2024 preliminary results of the LZ collaboration.} \cite{LZ:2022lsv, LZ:2024qwe} and PandaX-4T \cite{PandaX:2024qfu} through the Higgs or gauge portals, see the diagrams shown in Fig.~\ref{fig:ddetaxi}.

We performed a detailed numerical scan for the model parameters with various experimental and theoretical constraints. We have implemented the model in SARAH-4.15.1 and SPheno-4.0.5 \cite{Staub:2015kfa,Porod:2011nf} to calculate all the vertices, mass matrices and tadpole equations. The DM relic abundance, as well as the DM-nucleon scattering cross sections are determined by micrOMEGAS-5.3.41 \cite{Belanger:2014vza}. We have imposed the following conditions on all generated points:

\begin{itemize}
\item The parameters are taken in the ranges shown in Tab.~\ref{tab:cLFVparameter} and Tab.~\ref{tab:parameterrange}.
\item Bounded from below scalar potential, ensured by the vacuum stability constraints of Eq.~\ref{eq:vacuumstability}.
\item Perturbativity of Yukawas and quartic couplings as in Eqs.~\ref{eq:perturbativity1} and \ref{eq:perturbativity2}.
\item The DM is always ensured to be the lightest dark sector particle.
\item If $\eta^0$ is the DM particle, its mass must be smaller than the charged counterpart $\eta^+$. As can be seen from Eqs.~\ref{eq:metaplus} and \ref{eq:metamix}, this implies $\lambda_{5}<0$ which has been imposed for this case.
\item Neutrino oscillation parameters as in Sec.~\ref{sec:numass}.
\item We are applying the experimental constraints from the masses of the W and Z boson \cite{ParticleDataGroup:2020ssz}.
\end{itemize}

In addition, for all points satisfying correct relic abundance, we have also imposed the following constraints:
\begin{itemize}
\item We are imposing  the constraints from the oblique parameters at the $1\sigma$ level, as provided by the PDG \cite{ParticleDataGroup:2022pth}.
\begin{equation}
    S = -0.02 \pm 0.10, \,  T = 0.03 \pm 0.12,  \, U = 0.01 \pm 0.11.
\end{equation}
\item We are applying the experimental constraint based on the mass of the Higgs boson \cite{ParticleDataGroup:2022pth}.
\item We are imposing the bound on the branching ratio of Higgs invisible decay to the dark sector particles from the recent LHC data \cite{CMS:2023sdw}.
\item We impose the LEP constraint on the light-neutral component of a doublet. This limit is actually simply $m_{\eta R} + m_{\eta I} > m_Z$, which in our case translates to $m_{\eta^0} > m_Z/2 \approx 45.6$ GeV. In the case of the charged scalar component, this limit is $m_{\eta^+} \gsim 70 \text{  GeV}$ \cite{Belyaev:2016lok,Cao:2007rm}.
\item Finally, all blue points also satisfy the bounds from charged lepton flavor violation described in Sec.~\ref{sec:CLFV}.
\end{itemize}

\begin{table}
\begin{center}
\begin{tabular}{|    c   |    c    | c | c |}
  \hline 
  Parameter    &   Range   &   Parameter    &   Range  \\
\hline
$\lambda_{1}$     &  	 $[10^{-8},\sqrt{4\pi}]$            &  $\lambda_{2}$     &  	 $[10^{-8},\sqrt{4\pi}]$          \\
$\lambda_{3}$   &  $[10^{-8},\sqrt{4\pi}]$            &  $|\lambda_{4}|$     &      $[10^{-8},\sqrt{4\pi}]$          \\
$|\lambda_5|$   &   $[10^{-8},\sqrt{4\pi}]$     &  $|\lambda_{6}|$     &      $[10^{-8},\sqrt{4\pi}]$   \\
$|\lambda_{7}|$   &   $[10^{-8},\sqrt{4\pi}]$   	 &  $|\kappa|$          &	     $[10^{-8},30]\text{ GeV}$    \\
$\mu^2_{\eta}$  &  $[10^{2},10^{8}]\text{ GeV}^2$       &  $\mu^2_{\xi}$     & $[10^{2},10^{8}]\text{ GeV}^2$\\
$M_{N_1}$ & $[10,10^{4}] \text{ GeV}$ 	 & 
$M_{N_2}$ & $[10,10^{4}] \text{ GeV}$  \\	
    \hline
  \end{tabular}
\end{center}
\caption{Value range for the numerical parameter scan for relic density and DM direct detection. Other parameters are varied in accordance with Tab.~\ref{tab:cLFVparameter}.}
 \label{tab:parameterrange} 
\end{table}

These constraints are applied throughout the rest of the analysis wherever applicable. We now separately study the three possible cases regarding the nature of DM.

\subsection{Doublet Scalar Dark Matter}
\label{sec:doubletDM}
\begin{figure}[th]
 \centering
\hspace{-2cm} \textbf{\Large $\eta^0$ as DM}\\
        \includegraphics[height=6.5cm]{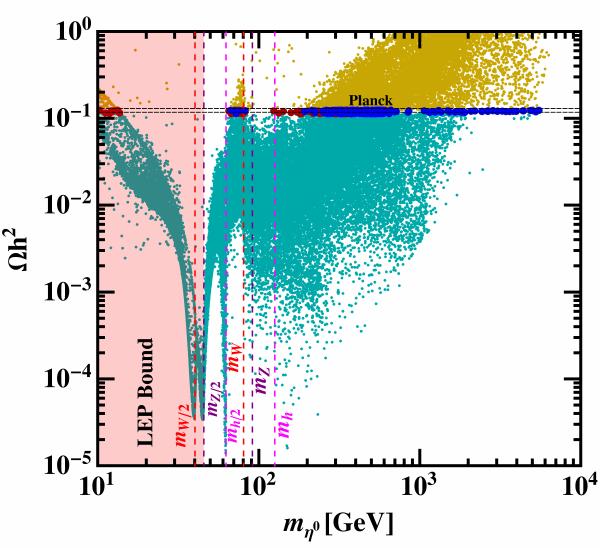}
        \includegraphics[height=6.5cm]{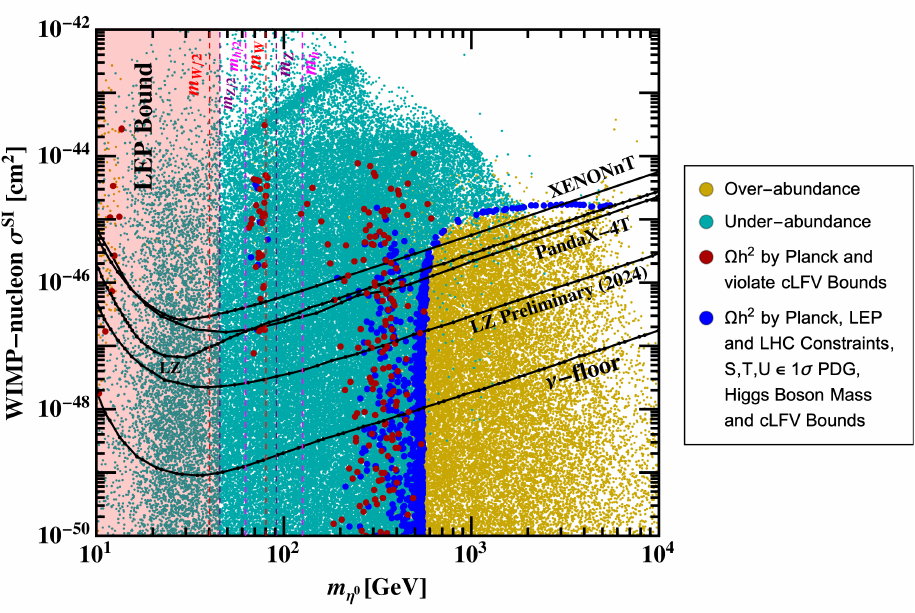}\\
        \caption{Predictions for the mainly doublet scalar DM case. In both panels, yellow/cyan points represent over/under abundant relic density \cite{Planck:2018vyg}, respectively. The red points satisfy the correct relic density but are ruled out by cLFV bounds. The blue points satisfy the correct relic density, LEP constraints, cLFV bounds, constraints from the oblique parameters, Higgs boson mass and Higgs invisible branching ratio at 95\% confidence level\cite{ATLAS:2022yvh}. \textbf{Left panel:} Relic density against mass of the doublet scalar DM particle. \\ \textbf{Right panel:} Spin-independent WIMP-nucleon direct detection cross section against mass of the doublet scalar DM particle. The exclusion limits from current generation DM direct detection experiments are shown by solid black lines. All the points above the exclusion lines are ruled out. }
        \label{fig:doubletDM}
\end{figure}

We start our DM analysis by focusing on the case in which the lightest particle of the dark sector is the doublet neutral scalar $\eta^0$. The left panel of Fig.~\ref{fig:doubletDM} shows the dependence of the DM relic abundance on the mass of the doublet scalar DM particle, taking into account its annihilation and co-annihilation diagrams into SM particles, shown in Fig~\ref{fig:anihieta} and Fig~\ref{fig:coanihieta}. 

In the relic density plot (left panel), we see the first three dips in the relic density when the DM mass is around half the mass of the $W$, $Z$ or Higgs bosons. At these masses, the annihilation of doublet scalar DM particles can occur very efficiently through the exchange of a $W$, $Z$ or Higgs bosons, respectively. As a result, the annihilation cross-section is enhanced, leading to a decrease in its relic density. The relevant mass scales have been marked in the plot as vertical lines.

The cyan/yellow points in the plot represent the points where $\eta^0$ is under-abundant/over-abundant. The under-abundance does not mean that $\eta^0$ is ruled out as a DM candidate, it merely means that it cannot be the sole DM candidate and the model needs to be extended to incorporate multi-component DM. Since such multi-component DM requires its own study, we do not go into details about this possibility. The over-abundance means that we have more DM compared to the Planck observations \cite{Planck:2018vyg}. 
The points satisfying the correct relic density are shown in two colors. The red colored points have correct relic density but are ruled out by cLFV constraints while the blue points satisfy the correct relic density, LEP constraints, cLFV bounds, constraints from the oblique parameters, Higgs boson mass and Higgs invisible branching ratio at 95\% confidence level\cite{ATLAS:2022yvh}. Note that the blue points occupy four different mass regions in the relic density plot: The first regime from 10 GeV to 15 GeV, the second regime from 65 GeV to 85 GeV, third regime from $\sim$ 200 GeV to $\sim$ 550 GeV and the fourth regime from $\sim$ 550 GeV to $\sim$ 5.4 TeV. 

The right panel of Fig.~\ref{fig:doubletDM} instead shows relationship between the spin-independent WIMP-nucleon cross-section through the Higgs and $Z$ portals, see Fig.~\ref{fig:ddetaxi} and the mass of the doublet scalar DM particle. The LZ experiment \cite{LZ:2022lsv, LZ:2024qwe} has established an upper bound on the nucleon cross-section, and this bound places constraints on WIMP masses over 6 GeV. The combination of all relevant constraints leads to an allowed mass range of $m_{\eta^0} \sim 70$ GeV $-$ 80 GeV and $m_{\eta^0} \sim 200$ GeV $-$ 600 GeV for mainly doublet scalar DM.  However, if we also take into account the recent preliminary results of LZ collaboration \cite{LZ:2024qwe}, then the allowed parameter space reduces to $m_{\eta^0} \sim 200$ GeV $-$ 600 GeV as shown in right panel of Fig.~\ref{fig:doubletDM}. We can divide the above analysis into Hierarchical and Co-annihilation regimes.

\subsubsection{Hierarchical Regime}

A blue region with DM in the high-mass from $\sim$ 550 GeV onwards is obtained when the mass of the doublet neutral scalar is significantly smaller than other dark sector particles i.e. $m_{\eta^0} \ll m_{\xi}, m_{N_1}$. 
\begin{figure}[th]
 \centering
 \textbf{\large Hierarchical ($\Delta m_{\xi\eta}, \Delta m_{N\eta} \gtrsim 20$ GeV)}\\
        \includegraphics[height=6.5cm]{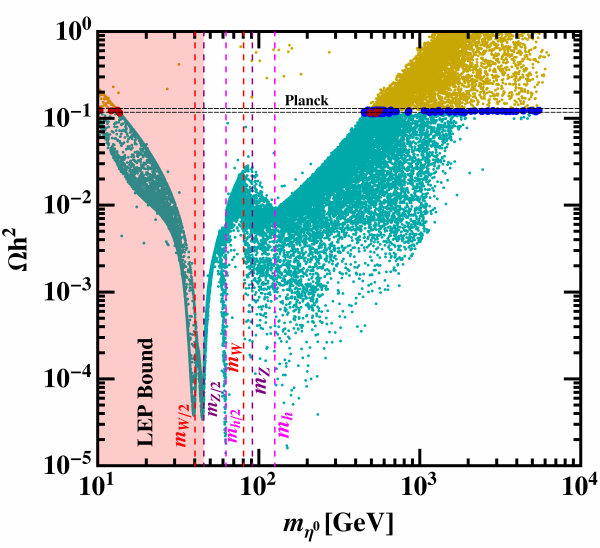}
        \includegraphics[height=6.5cm]{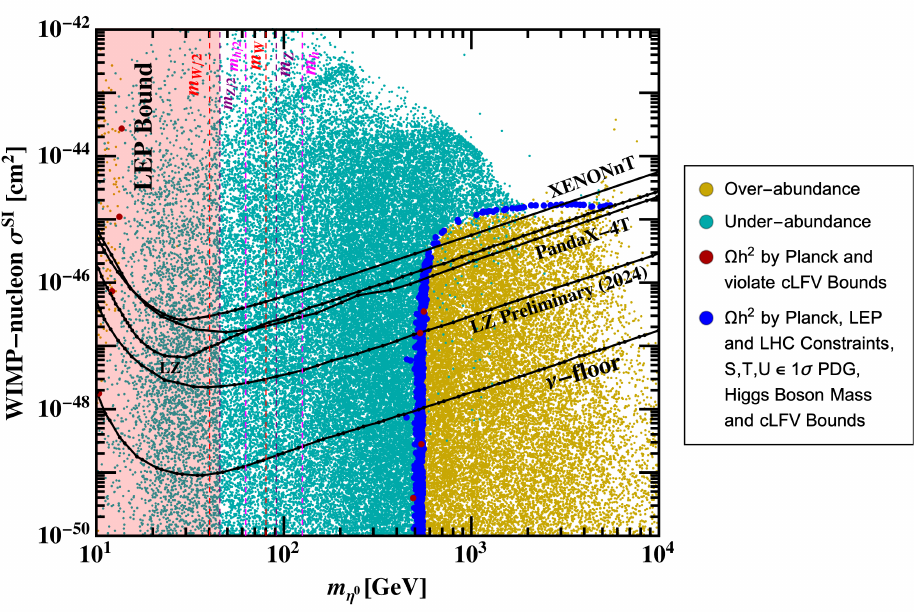}\\
        \caption{Predictions for the doublet scalar DM case when the particles in the dark sector are hierarchical, i.e. both $ \Delta m_{\xi\eta} \,$ and $\Delta m_{N\eta} \, \gtrsim 20$ GeV. Other details are same as in Fig.~\ref{fig:doubletDM}.}
        \label{fig:doubletDMhierarchichal}
\end{figure}
Note that, in this regime, both the relic density and the DM annihilation cross-section are determined predominantly by the $H-\eta$ couplings $\lambda_{4}$ and $\lambda_{5}$, as well as the known gauge couplings and the DM mass.
The lower mass region is ruled out by aforementioned LEP constraints. There are also red points in the lower mass region so it is also ruled out by cLFV bounds. In the middle mass region, there are no points that satisfy the correct relic density as there is efficient annihilation of DM through the exchange of h, $W$ and Z bosons. In the high-mass region, we have only a few red points but most of the points are blue i.e. they satisfy all constraints including cLFV. Therefore, cLFV bounds do not significantly constrain the high-mass region. 
Note that in the right panel of Fig. \ref{fig:doubletDMhierarchichal}, we  observe a prominent vertical line at approximately 550 GeV. This is because of the parameter space opening up for DM that satisfies the correct relic density. Here, the couplings $\lambda_{4}$ and $\lambda_{5}$ primarily regulate the DM annihilation cross-section, while the relic density can be kept at the desired value by adjusting other parameters,  leading to blue points satisfying both relic abundance and direct detection constraints.

As the DM mass exceeds beyond approximately 5.4 TeV, loop corrections to the Higgs mass become too large to balance with the tree-level Higgs-quartic ($\lambda_1$) coupling. Therefore, the blue points stop at around 5.4 TeV in the relic density plot. The combination of all relevant constraints leads to an allowed mass range of $m_{\eta^0} \sim 500 - 600$ GeV for this scenario which is similar to that obtained in previous works \cite{CentellesChulia:2022vpz,CentellesChulia:2024svj}.

\subsubsection{Co-annihilation Regime}

\begin{figure}[th]
 \centering
  \textbf{\large Co-annihilation 1 ($\Delta m_{\xi\eta} \lesssim 20$ GeV)}\\
        \includegraphics[height=6.5cm]{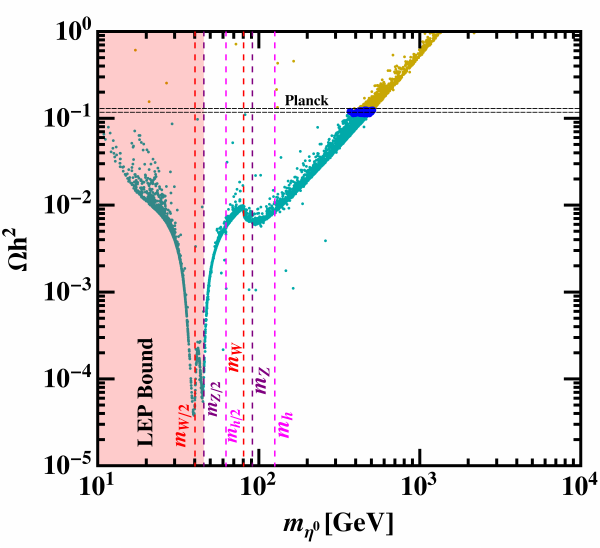}
        \includegraphics[height=6.5cm]{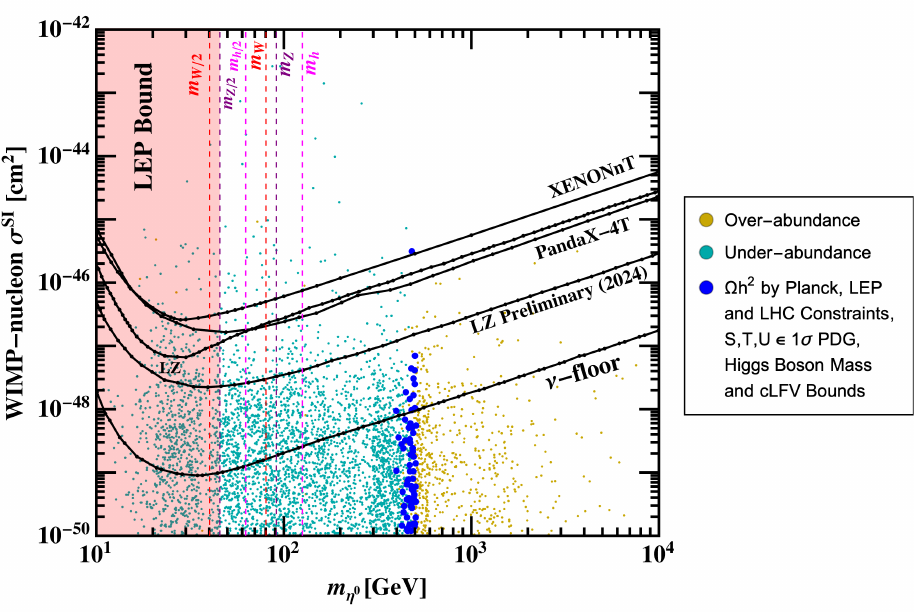}\\
          \textbf{\large Co-annihilation 2 ($\Delta m_{N\eta} \lesssim 20$ GeV)}\\
        \includegraphics[height=6.5cm]{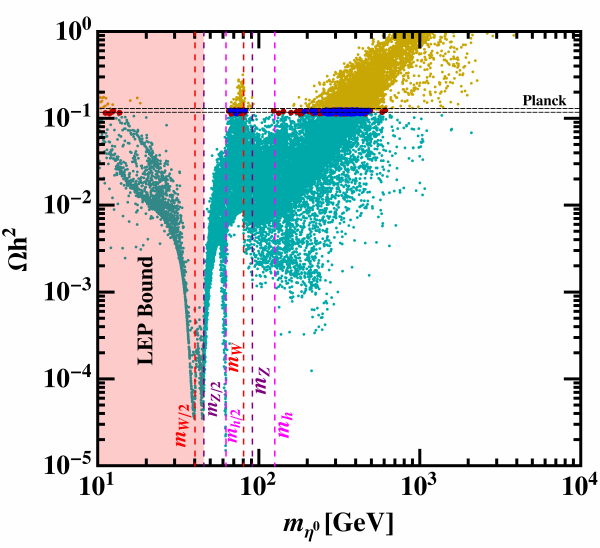}
        \includegraphics[height=6.5cm]{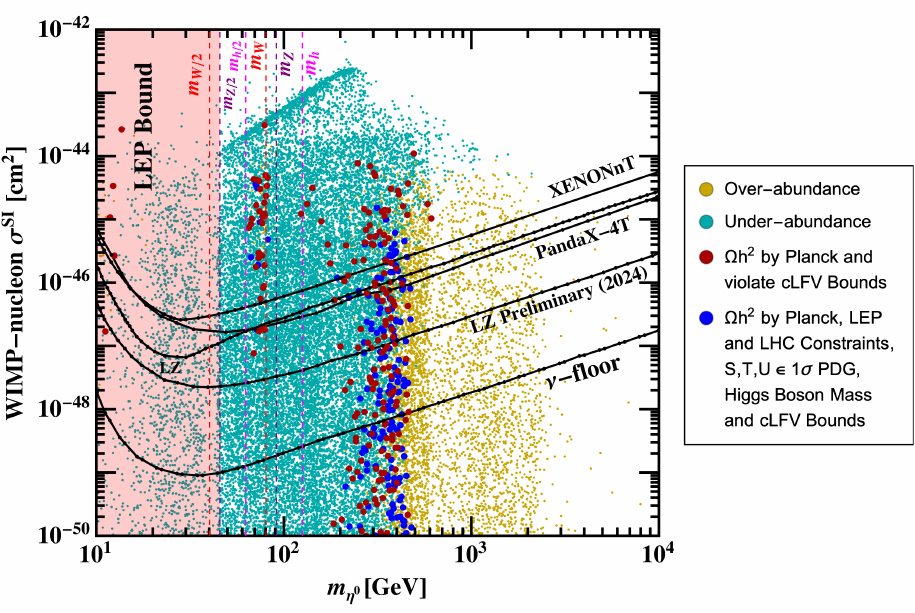}\\
        \caption{Predictions for the doublet scalar DM case in the co-annihilation regime. 
    Upper Panel (co-annihilation 1) is for the case $ \Delta m_{\xi\eta}  \lesssim 20$ GeV  while lower panel (co-annihilation 2)  is for the case $\Delta m_{N\eta} \lesssim 20$ GeV.  Other details are same as in Fig.~\ref{fig:doubletDM}.}
        \label{fig:doubletDMcoannihilation}
\end{figure}

Alternatively, the masses of the other particles in the dark sector may be close to that of $\eta^0$. If this is the case, new co-annihilation diagrams become important, see Fig.~\ref{fig:coanihieta}. There are two possible co-annihilation scenarios occurs when: 1) $\Delta m_{\xi\eta} \, = \, m_\xi \, - \, m_{\eta^0} \lesssim 20$ GeV and 2) $\Delta m_{N\eta} \, = \,  m_{N_1} \, - \, m_{\eta^0} \lesssim 20$ GeV. The results for both cases are shown in Fig. \ref{fig:doubletDMcoannihilation}. 

The upper panel in Fig.~\ref{fig:doubletDMcoannihilation} is for the case when $\Delta m_{\xi\eta} \lesssim 20$ GeV while the lower panel is for the case when $\Delta m_{N\eta} \lesssim 20$ GeV.
In the first co-annihilation scenario ($\Delta m_{\xi\eta} \lesssim 20$ GeV), additional co-annihilation channels involving singlet scalar interactions appear. In this case, we obtain a new parameter space in the medium mass region due to co-annihilation effects between singlet and neutral doublet scalar as shown by blue points. There are no red points for this regime as the mass of singlet fermion is comparatively higher for correct relic points leading to lower cLFV rates.
 
In the second co-annihilation scenario ($\Delta m_{N\eta} \lesssim 20$ GeV), additional co-annihilation channels involving dark fermion interactions appear. Here, the fermion mass is comparable to the mass of the doublet neutral scalar, which enhances the co-annihilation effect.  This leads to a new parameter space in the medium mass region as shown by blue points. The lower mass region is excluded by cLFV bounds and the previously mentioned LEP constraints. While a few red points are found in the medium mass range, blue points are also present. Consequently, two mass regions $m_{\eta^0} \sim 70$ GeV $-$ 80 GeV and $m_{\eta^0} \sim 200$ GeV $-$ 600 GeV emerge as the viable regions that satisfy all constraints. However, if we also take into account the recent preliminary results of LZ collaboration \cite{LZ:2024qwe}, then the allowed parameter space reduces to $m_{\eta^0} \sim 200$ GeV $-$ 600 GeV as shown in the lower right panel of Fig.~\ref{fig:doubletDMcoannihilation}.

It is important to mention that the results for the doublet scalar DM case are equivalent to those in the Majorana scotogenic case \cite{Batra:2022pej}. However, in the scotogenic Dirac case, there is an alternative for having scalar DM, namely the singlet scalar DM which we discuss in Sec.~\ref{sec:singletDM}.

\FloatBarrier
\subsection{Singlet Scalar Dark Matter}
\label{sec:singletDM}

\begin{figure}[th]
 \centering
\hspace{-2cm} \textbf{\Large $\xi$ as DM}\\
        \includegraphics[height=6.5cm]{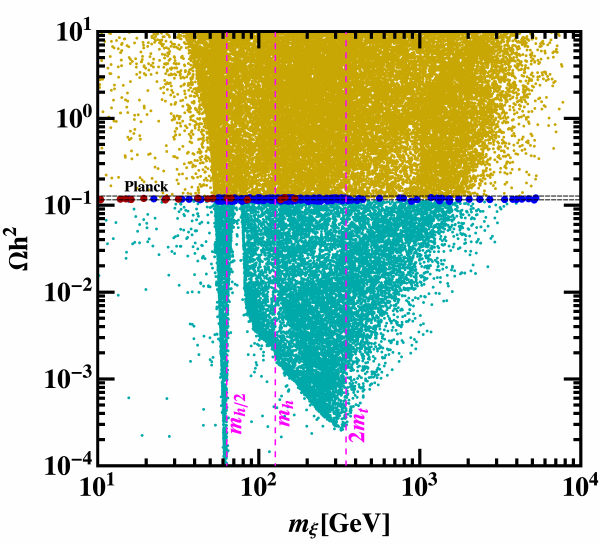}
        \includegraphics[height=6.5cm]{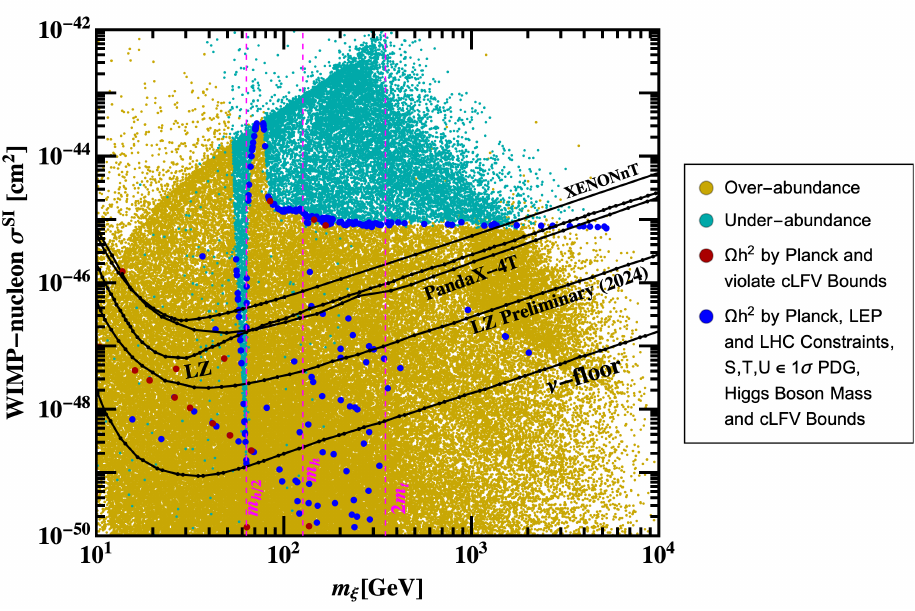}\\
         \caption{Predictions for the mainly singlet scalar DM case. The color code is the same as in Fig.~\ref{fig:doubletDM}. \textbf{Left panel:} Relic density against mass of the singlet scalar DM particle. \textbf{Right panel:} Spin-independent WIMP-nucleon direct detection cross section against mass of the singlet scalar DM particle. Note that in this case, due to co-annihilation of DM with other dark sector particles, the DM can be quite light.}
        \label{fig:singletDM}
\end{figure}

We now move on to the case in which the DM particle is the $SU(2)_L$ singlet scalar $\xi$. If this field is lighter than the neutral and charged doublet components as well as the neutral fermion i.e. $m_\xi < m_N, m_{\eta^0}, m_{\eta^+}$, it will be the DM particle of the model. The left panel of Fig.~\ref{fig:singletDM} 
shows the dependence of the DM relic abundance on the mass of the singlet scalar DM particle. The relic density is computed by taking into account its annihilation and co-annihilation diagrams into bosons and fermions, shown in Fig.~\ref{fig:anihixi} and~\ref{fig:coanihixi}. The color code is the same as in the  
previous section. One, however, has to keep in mind that for the current case, the dominant DM annihilation channel is only the Higgs portal as shown in Fig.~\ref{fig:anihixi} while the WIMP-nucleon cross-section is also mostly controlled through the Higgs mediation as shown in Fig.~\ref{fig:ddetaxi}. We see a sharp dip in the relic density when the DM mass is around half the mass of the Higgs boson. At this mass, the annihilation of singlet scalar DM particles can occur very efficiently through the exchange of a Higgs boson. As a result, the annihilation cross-section is enhanced, leading to a decrease in its relic density. When the DM mass is roughly twice the top quark's mass, annihilation of DM via the Higgs boson into top quark pairs becomes kinematically allowed. Around this threshold, the annihilation cross-section increases significantly, leading to a more efficient depletion of DM particles. This can be observed as the final dip in the relic density plot.

The right panel of Fig.~\ref{fig:singletDM} shows the relationship between spin-independent WIMP-nucleon cross-section and the mass of the singlet scalar DM particle through the Higgs portal, see Fig.~\ref{fig:ddetaxi}. There are a few red points in the lower and medium mass region but a significant number of blue points as well. The blue points cover a wide region starting from 10 GeV to around 5.2 TeV in the relic density plot. As the DM mass exceeds approximately 5.2 TeV, loop corrections to the Higgs mass become too large and cannot be counterbalanced by decreasing the tree-level Higgs-quartic ($\lambda_1$) coupling. Therefore, these blue points stop at around 5.2 TeV in Fig.~\ref{fig:singletDM}. The combination of all relevant constraints leads to an allowed mass range of $m_{\xi} \sim 10$ GeV $-$ 2 TeV (co-annihilation) and $m_{\xi} \sim 3.5$ TeV $-$ 5.2 TeV (hierarchical) for singlet scalar DM. However, if we also take into account the latest preliminary results of LZ collaboration \cite{LZ:2024qwe} then the allowed parameter space reduces to $m_{\xi} \sim 10$ GeV $-$ 2 TeV as the even higher mass region also gets ruled out.
Similar to the previous case, we can divide the above analysis into hierarchical and co-annihilation regimes.

\subsubsection{Hierarchical Regime}

\begin{figure}[th]
\centering
\textbf{\large Hierarchical ($\Delta m_{\eta\xi}, \Delta m_{N\xi} \gtrsim 100$ GeV)}\\
        \includegraphics[height=6.5cm]{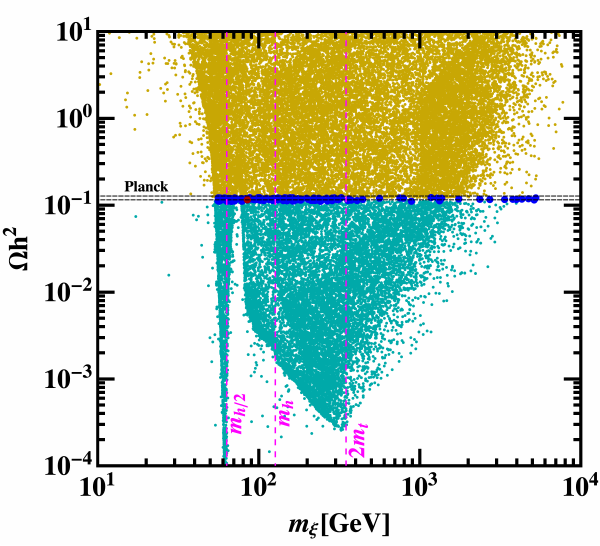}
        \includegraphics[height=6.5cm]{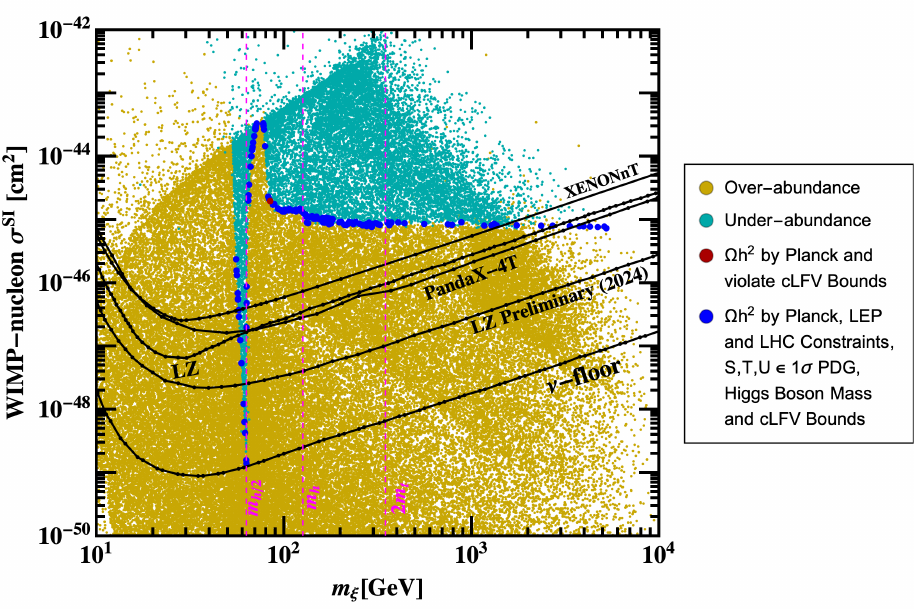} \\
        \caption{Predictions for the singlet scalar DM case in the hierarchical regime, i.e. both $\Delta m_{\eta\xi} \,$ and $\Delta m_{N\xi} \, \gtrsim 100$ GeV. Other details are same as in Fig.~\ref{fig:doubletDM}.}
        \label{fig:singletDMhierarchical}
\end{figure}

The first one, shown in Fig.~\ref{fig:singletDMhierarchical}, corresponds to the case where the singlet scalar mass is not close to the mass of other dark sector particles, i.e. $ \Delta m_{\eta\xi} = m_{\eta^0} - m_{\xi}$, $\Delta m_{N\xi} = m_{N_1} - m_{\xi}\gtrsim 100$ GeV.    
The left panel of Fig.~\ref{fig:singletDMhierarchical} shows the dependence of DM relic abundance on the mass of singlet scalar DM particle, taking into account its annihilation diagrams into bosons and fermions, shown in Fig~\ref{fig:anihixi}. In this scenario, the main contribution to both the DM annihilation and direct detection is through the Higgs portal, i.e. the term $\lambda_6(H^{\dagger} H \xi^{*} \xi$) of the scalar potential given in Eq.~\ref{eq:pot}.

The right panel of Fig.~\ref{fig:singletDMhierarchical} instead shows relationship between the spin-independent WIMP-nucleon cross-section and the mass of the singlet scalar DM particle through the Higgs portal, see Fig.~\ref{fig:ddetaxi}. Note that, in this regime, both the relic density and the DM annihilation cross-section are determined only by the $H-\xi$ couplings $\lambda_{6}$, as well as the known gauge couplings and the DM mass. Therefore, there is a strong correlation between relic density and direct detection cross section as shown in the right panel of Fig.~\ref{fig:singletDMhierarchical}. There is only a single red point at around $W$ boson mass but there are also many blue points. Therefore, cLFV bounds do not have such role for this regime. As the DM mass exceeds approximately 5.2 TeV, loop corrections to the Higgs mass become too strong and cannot be counterbalanced by adjusting the tree-level Higgs-quartic $\lambda_1$ coupling. The combination of all relevant constraints leads to a very tiny mass range at around half Higgs mass and above 3.5 TeV mass. However, if we also take into account the recent preliminary results of LZ collaboration \cite{LZ:2024qwe}, then the high-mass range gets completely ruled out as shown in right panel of Fig.~\ref{fig:singletDMhierarchical}.

\subsubsection{Co-annihilation Regime}

\begin{figure}[th]
\centering
\textbf{\large Co-annihilation 1 ($\Delta m_{\eta\xi} \lesssim 100$ GeV)}\\
        \includegraphics[height=6.5cm]{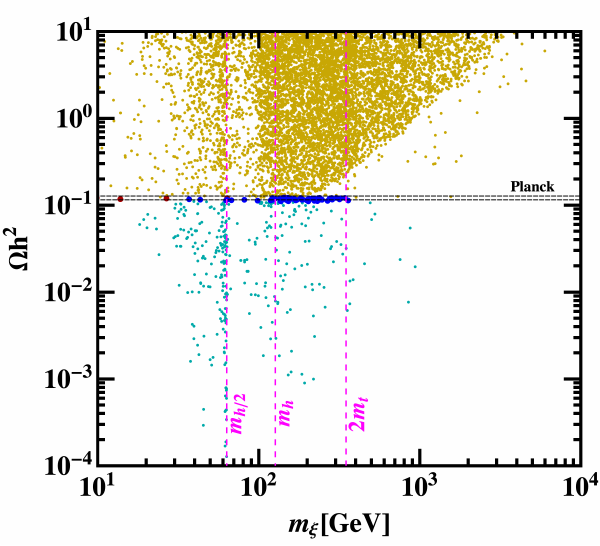}
        \includegraphics[height=6.5cm]{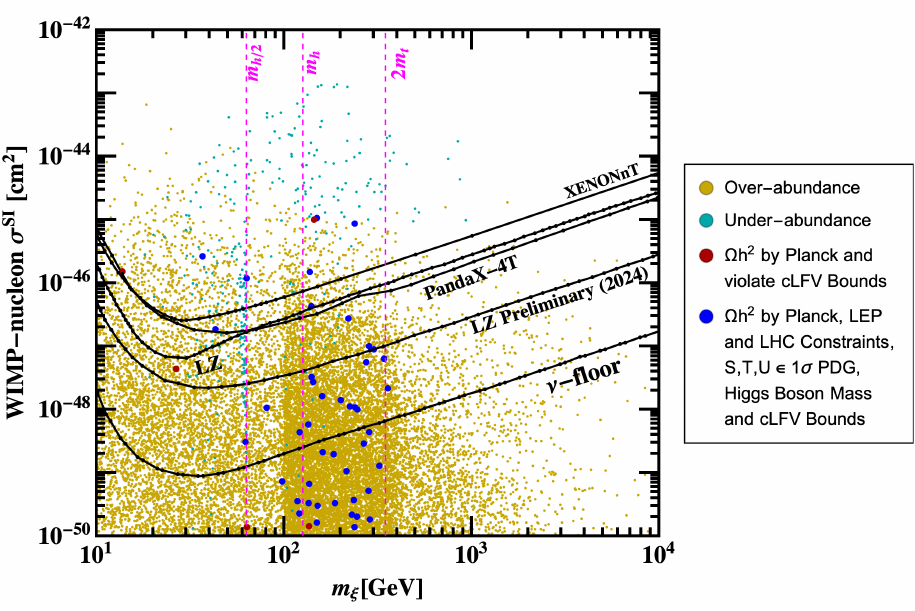} \\
\textbf{\large Co-annihilation 2 ($\Delta m_{N\xi} \lesssim 100$ GeV)}\\
        \includegraphics[height=6.5cm]{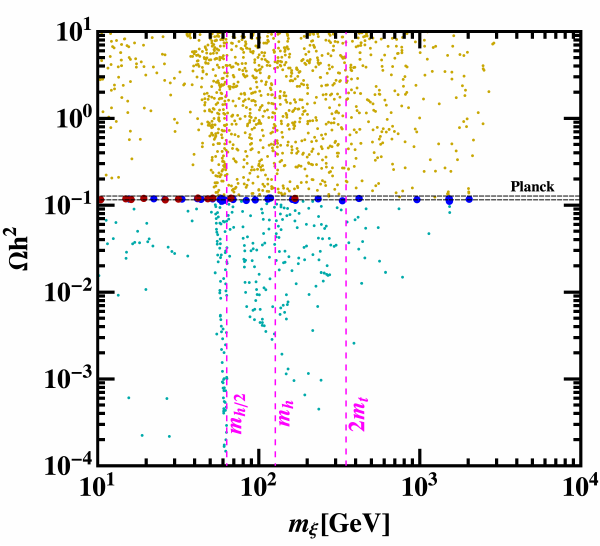}
        \includegraphics[height=6.5cm]{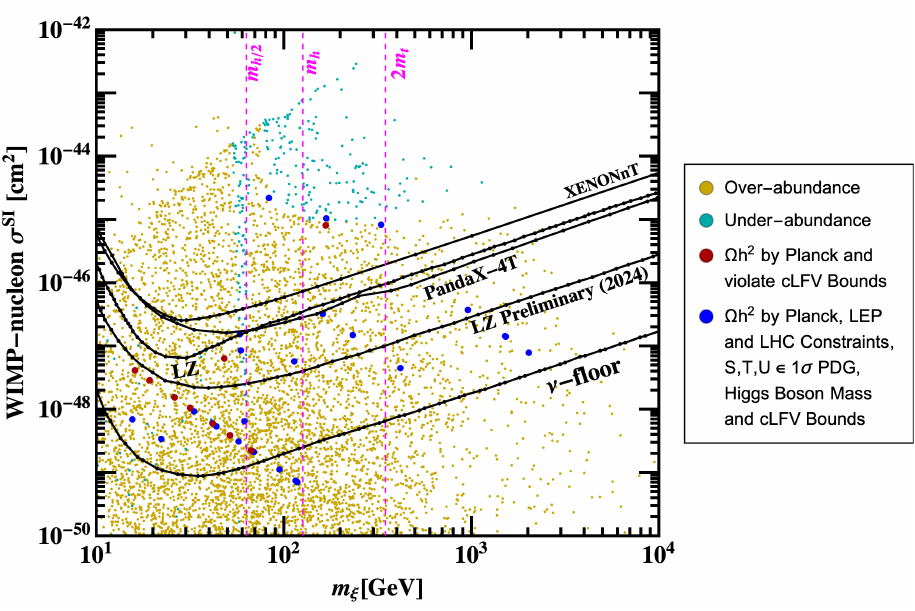} \\
        \caption{Predictions for the singlet scalar DM case in the co-annihilation regime.
        Upper panel (co-annihilation 1) is for $\Delta m_{\eta\xi} \lesssim 100$ GeV and lower panel (co-annihilation 2) is for $\Delta m_{N\xi} \lesssim 100$ GeV .  Other details are same as in Fig.~\ref{fig:doubletDM}.} 
        \label{fig:singletDMcoannihilation}
\end{figure}

Alternatively, the masses of the other particles in the dark sector may be close to that of $\xi$. If this is the case, new co-annihilation diagrams become important, see Fig.~\ref{fig:coanihixi}. There are two possible co-annihilation scenarios when: 1) $\Delta m_{\eta\xi} \lesssim 100$ GeV and 2) $\Delta m_{N\xi} \lesssim 100 $ GeV. The results for both cases are shown in Fig. \ref{fig:singletDMcoannihilation}.

In the first co-annihilation scenario ($\Delta m_{\eta\xi} \lesssim 100$ GeV), additional co-annihilation channels involving neutral component of doublet scalar appear. Both the relic density and the direct detection cross section now also depend on the $Y, Y'$ Yukawa couplings, see Fig.~\ref{fig:coanihixi}. Therefore, the correlation between these two observables is broken in this scenario. We obtain the correct relic points in the medium mass region as shown in the upper panel of Fig.~\ref{fig:singletDMcoannihilation}. We have a few red points in the lower mass region i.e., ruled out by cLFV bounds. The lower mass region is also ruled out by aforementioned LEP constraints. Therefore, the only surviving region is the medium mass region for this scenario.

Again, in the second co-annihilation scenario ($\Delta m_{N\xi} \lesssim 100 $ GeV), additional co-annihilation channels involving dark fermion interactions appear and Yukawa coupling $Y^{\prime}$ plays an important role, see Fig.~\ref{fig:coanihixi}.
In our analysis, for singlet scalar DM mass between 10 GeV to 2 TeV, the blue points satisfy all the aforementioned constraints, while the red points get ruled out by cLFV. Exploration into higher mass regions results in an increase in relic density values. Therefore, to maintain correct relic density values, it becomes necessary to increase the values of $Y^\prime$ Yukawa couplings.  In the scanned region, the parameter space terminates at around 2 TeV after applying all the relevant constraints.\\
It is important to note that for the $\Delta m_{N\xi}$ case, the co-annihilation is happening between SM gauge singlet dark sector particles. Therefore, the aforementioned constraints from LEP are not applicable in the lower mass region. This is because the mass of doublet scalar components, which in this case can be easily taken heavy, well beyond LEP limits. 
However, in this region the cLFV rates can be high leading to many red points in lower panel of Fig.~\ref{fig:singletDMcoannihilation}. This happens because the dark fermion is close in mass to DM and the charged scalar $\eta^\pm$ mass can also be low, being just outside the LEP limit. 
However, this low DM mass region also has many blue points corresponding to parameter space where the $Y$ Yukawa couplings are small and/or $\eta^\pm$ mass is very large. Since, in this case neither $Y$ nor $\eta^\pm$ mass plays any role in satisfying relic density, they can be taken arbitrarily small and large, respectively, leading to the blue points.
Hence, unlike the simple Higgs portal DM models \cite{Cline:2013gha, Wu:2016mbe} or even the Majorana Scotogenic models \cite{Batra:2022pej}, the DM in Dirac Scotogenic model can be of very low-mass. This is a unique feature of Dirac Scotogenic model.

\FloatBarrier
\subsection{Fermionic Dark Matter}
\label{sec:fermionDM}

\begin{figure}[th]
 \centering
 \textbf{\large \hspace{-2cm} Fermionic DM}\\
        \includegraphics[height=7cm]{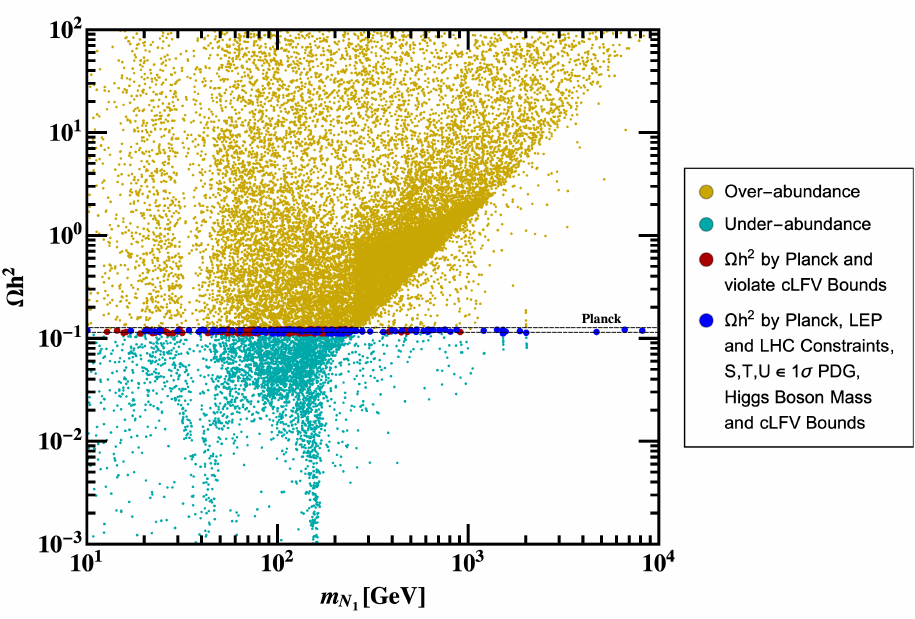}\\
        \caption{Relic density against mass of the fermion DM particle. The color code is the same as in Fig.~\ref{fig:doubletDM}. }
        \label{fig:fermionDM}
\end{figure}

Finally, we consider the possibility of fermionic DM, i.e., the scenario where the lightest dark sector particle is a fermion. To be definite, we choose $N_1$ as the lightest dark sector particle. This implies that  $m_{N_1} < m_{N_2} $ and it is lighter than both the neutral and charged component of the doublet scalar ($m_{N_1} < m_{\eta^0}, m_{\eta^+}$) as well as the singlet scalar ($m_{N_1} < m_\xi$).
Since the $N_1$ has a gauge invariant mass, we can easily take it to be lighter than other dark sector particles. Fig.~\ref{fig:fermionDM} shows the dependence of the DM relic abundance on the mass of the fermion DM particle. 

As clear from Fig.~\ref{fig:fermionDM}, the relic density values are exceptionally high for most of the points in this case.
To achieve the correct relic density, it becomes crucial to invoke the co-annihilation channels given in Fig.~\ref{fig:coanihiN}. As can be seen from Fig.~\ref{fig:fermionDM}, taking into account the co-annihilation channels allows us to obtain the correct relic density.
In Fig.~\ref{fig:fermionDM}, there are a few red points in the lower and medium mass region but a significant number of blue points as well. Importantly, being fermionic DM, the constraints from direct detection experiments such as XENONnT \cite{XENON:2023cxc}, LZ \cite{LZ:2022lsv, LZ:2024qwe}, and PandaX-4T \cite{PandaX:2024qfu} are not significant.
We obtain a wide range of parameter space that satisfy all previously discussed constraints by the inclusion of co-annihilation channels as shown in Fig.~\ref{fig:fermionDM}. To highlight the crucial role of co-annihilation, we can divide the above analysis into Hierarchical and Co-annihilation regimes.

\FloatBarrier

\subsubsection{Hierarchical Regime}
\begin{figure}[th]
\centering   
\textbf{\large Hierarchical ($\Delta m_{\eta N}, \Delta m_{\xi N} \gtrsim 100$ GeV)}\\
        \includegraphics[height=7cm]{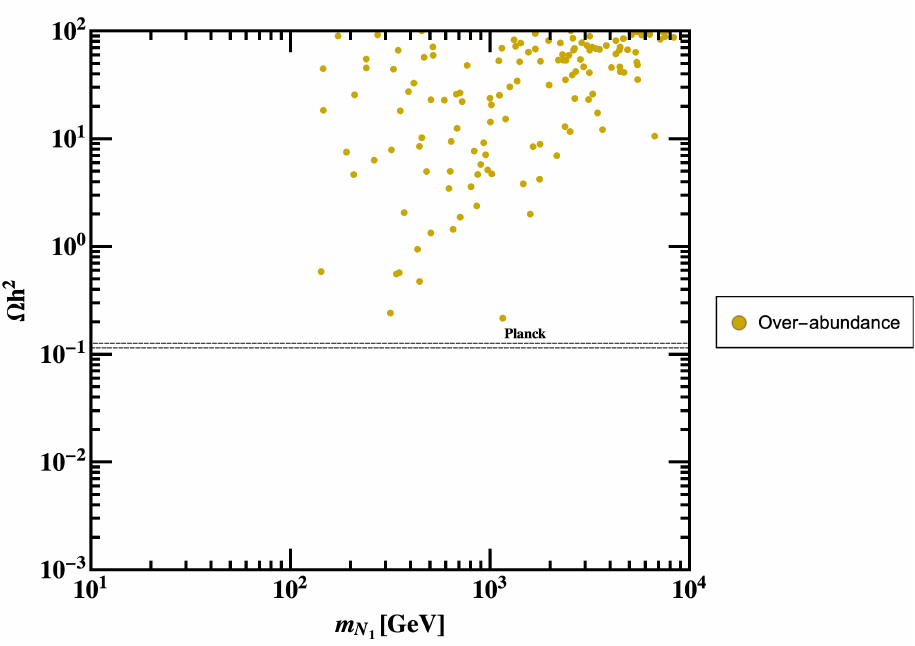}\\
        \caption{Predictions for the fermion DM case in the hierarchical regime, i.e. both $\Delta m_{\eta N} \,$ and $\Delta m_{\xi N} \, \gtrsim 100$ GeV. Other details are same as in Fig.~\ref{fig:fermionDM}.}
        \label{fig:fermionDMhierarchical}
\end{figure}

In the hierarchical regime, the mass of the dark fermion is significantly smaller than that of the neutral doublet scalar and singlet scalar i.e. $\Delta m_{\eta N} \equiv m_{\eta^0}-m_{N_1} \gtrsim 100 $ GeV and  $\Delta m_{\xi N} \equiv m_{\xi}-m_{N_1} \gtrsim 100 $ GeV. 
When both $\Delta m_{\eta N}, \Delta m_{\xi N} \gtrsim 100$ GeV, co-annihilation channels do not contribute significantly to the computation of relic density. Instead, the relic density is predominantly determined by fermion annihilation channels shown in Fig.~\ref{fig:anihiN}, resulting in a higher relic density in this regime as shown in Fig.~\ref{fig:fermionDMhierarchical}.

In such a case, no point that satisfies the correct relic density can be obtained \cite{Guo:2020qin}. Thus, for the fermionic DM case, co-annihilation channels play a crucial role in obtaining the correct relic density, as we discuss next.
\FloatBarrier

\subsubsection{Co-annihilation Regime}

\begin{figure}[th]
\centering   
\textbf{\large Co-annihilation 1 ($\Delta m_{\eta N} \lesssim 100$ GeV)}\\
        \includegraphics[height=7cm]{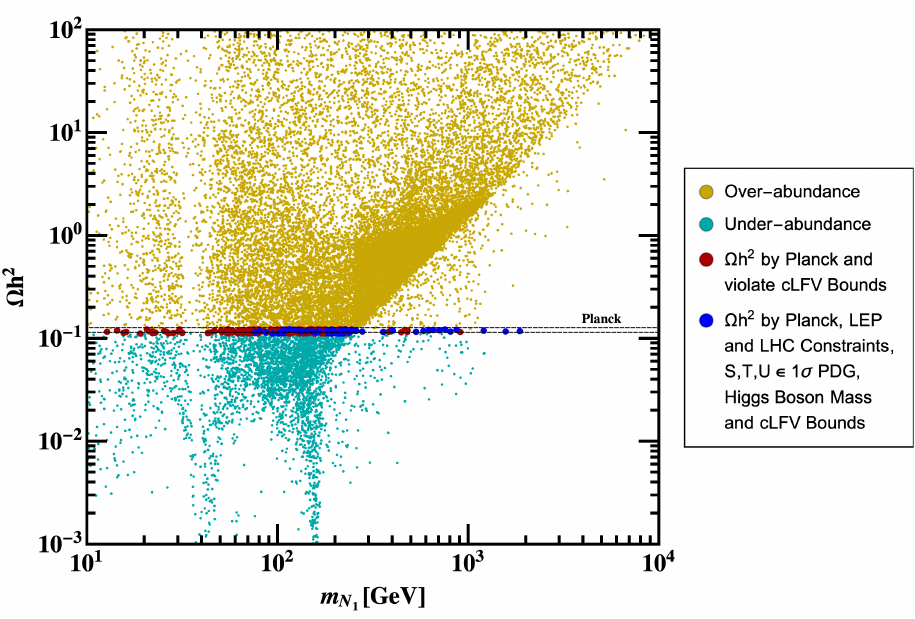}\\
\textbf{\large Co-annihilation 2 ($\Delta m_{\xi N} \lesssim 100$ GeV)}\\
        \includegraphics[height=7cm]{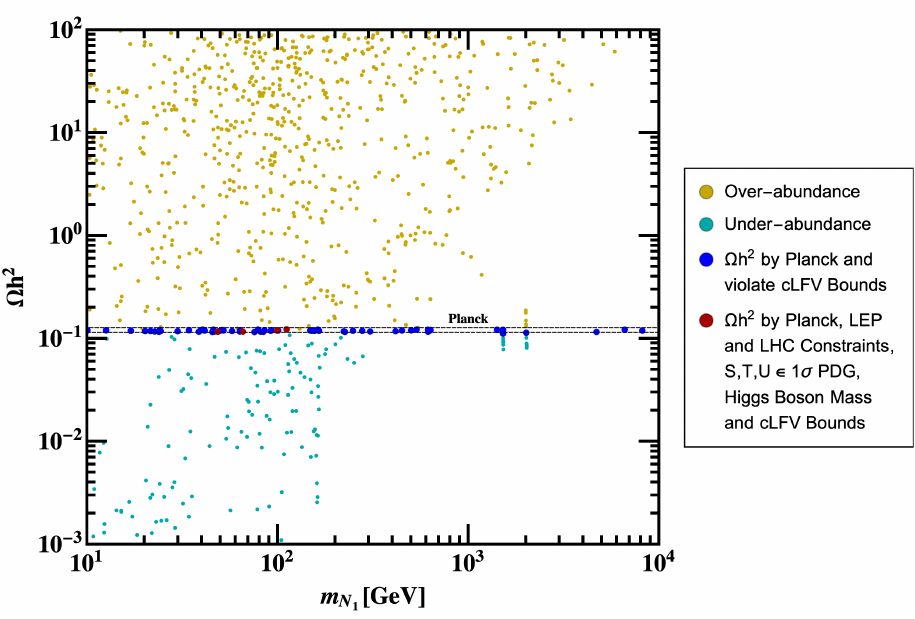}\\
        \caption{Predictions for the fermion DM case in the co-annihilation regime, Upper panel (co-annihilation 1) is for $\Delta m_{\eta N} \lesssim 100$ GeV and lower panel (co-annihilation 2) is for $\Delta m_{\xi N} \lesssim 100$ GeV.  Other details are same as in Fig.~\ref{fig:doubletDM}.}
        \label{fig:fermionDMcoannihilation}
\end{figure}

Alternatively, the masses of the other particles in the dark sector may be close to that of $N_{1}$. If this is the case, new co-annihilation diagrams become important, see Fig.~\ref{fig:coanihiN}. There are two possible co-annihilation scenarios when: 1) $\Delta m_{\eta N} \lesssim 100 $ GeV and 2) $\Delta m_{\xi N} \lesssim 100 $ GeV. The results for both cases are shown in Fig. \ref{fig:fermionDMcoannihilation}.

In the first co-annihilation scenario ($\Delta m_{\eta N} \lesssim 100 $ GeV), additional co-annihilation channels involving neutral doublet scalar interactions appear. In this regime, the lower mass region is excluded due to the LEP constraints as mentioned earlier. Additionally, this region is ruled out by cLFV bounds, as indicated by the numerous red points in the lower mass range. In the medium mass region, although a few red points appear, blue points are also present. Therefore, within the explored parameter space, the fermionic DM candidate $N_{1}$, the mass range from approximately 80 GeV to around 2 TeV, consistently satisfies all previously discussed constraints.

In the second co-annihilation scenario ($\Delta m_{\xi N} \lesssim 100 $ GeV), additional co-annihilation channels involving singlet scalar interactions appear. This regime becomes particularly relevant with higher values of the $Y^{\prime}$ Yukawa couplings, especially when the mass of the singlet scalar is close to that of the fermion DM candidate $N_{1}$. In this scenario, we achieve correct relic density starting from DM masses around 10 GeV. There are a few red points around 100 GeV but there are also many blue points so that region is unaffected by cLFV bounds. In the lower mass region, aforementioned constraints from LEP are not applicable due to much higher masses of the doublet scalar components compared to the dark fermion. However, this region typically exhibits high cLFV rates. These rates can be effectively reduced by decreasing the $Y$ Yukawa couplings and increasing the mass of the charged component of the doublet scalar. Despite these adjustments, the relic density remains largely unaffected because it primarily arises from the co-annihilation channels involving the singlet scalar.
Exploration into higher mass regions results in an increase in relic density values. Therefore, to maintain correct relic density values, it becomes necessary to increase the values of $Y^{\prime}$ Yukawa couplings. In the scanned region, the parameter space terminates at around 8.1 TeV after applying all the relevant constraints.
Therefore, within the explored parameter space, the fermionic DM candidate $N_{1}$, spanning a mass range from approximately 10 GeV to around 8.1 TeV, consistently satisfies all previously discussed constraints for this second regime case. 

In conclusion, for the fermionic DM case, co-annihilation involving dark scalars plays a crucial role, providing a wide range of viable parameter space. Importantly, the low-mass region is allowed due to co-annihilation channels involving the singlet scalar, a distinctive feature of the Dirac Scotogenic model which distinguishes it from the canonical Majorana Scotogenic model \cite{Batra:2022pej}.
\FloatBarrier

\section{Conclusions}
\label{sec:conclusions}
We have thoroughly analysed the phenomenology of the Dirac scotogenic model, including neutrino masses and mixing, DM relic abundance, stability of the electroweak vacuum, charged lepton flavor violation, and collider constraints.
When the DM is either the doublet scalar or the fermion, the cLFV rates may fall within the sensitivity of future experimental searches, offering promising prospects for detection. An important feature of the model is that, compared to the Majorana scotogenic case, new DM mass ranges are feasible due to the presence of the singlet scalar state and its contribution through co-annihilation channels. 

For the doublet scalar DM case two distinct mass regions are viable: a low-mass region near half the Higgs mass and a medium mass region just above the top quark mass. In the singlet scalar DM case, if we only consider the annihilation channels, the DM mass is restricted to either around half the Higgs mass or heavy enough ($\gtrsim 3.5$ TeV) that it escapes from direct detection bounds from current experiments. If we take the preliminary 2024 LZ results also into account then this heavy mass region will also be ruled out.
However, an alternative exists if the mass of the singlet scalar DM is close to the mass of the doublet scalar or fermion, in which case the co-annihilation channels become important, opening up a wide low-mass parameter space. This is a unique feature of the Dirac Scotogenic model not shared by other simple Higgs portal scalar DM models. Finally, we have also studied the fermionic DM case. Again, in this case, a wide range of viable parameter space is allowed by considering co-annihilation channels involving dark scalars.
  

\FloatBarrier
\newpage
\appendix

\section{cLFV Calculation in Dirac Scotogenic Model}
\label{sec:appendix2}

\begin{figure}[h!]
\includegraphics[width=8cm]{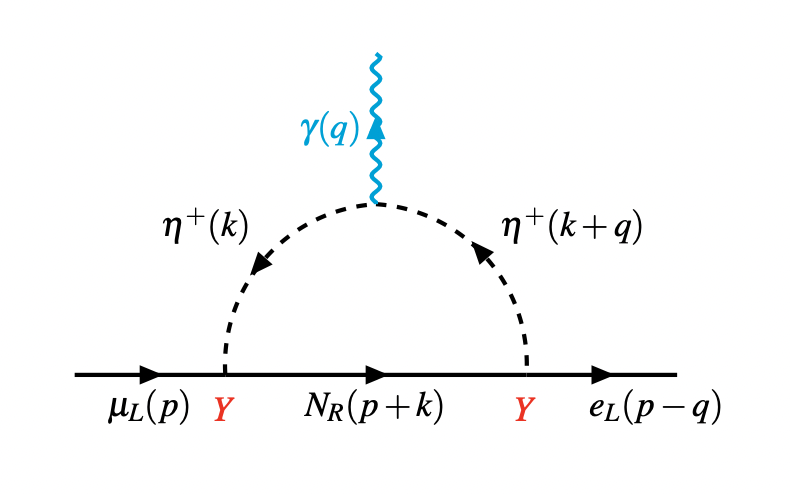}
        \caption{One-loop Feynman diagram for the process $\mu \rightarrow e \gamma$}
        \label{fig:LFV2}
\end{figure}

In the Dirac Scotogenic model, the main contribution to the cLFV decay $\mu \rightarrow e \gamma$ arises from the diagram in Fig.~\ref{fig:LFV2}. Its invariant amplitude $\mathcal{M}$ and the resulting branching ratio is calculated in this chapter. The arrows show the direction of the charge flow.

The invariant amplitude is compiled by using the Feynman rules

\begin{eqnarray}
\mathrm{i} \mathcal{M} & = & \sum_{i} \int_{\mathbb{R}^4} \frac{\mathrm{d}^4 k}{(2 \pi)^4} \bar{u_{e}}\left(p-q\right) Y_{e i}^{*}\frac{\left(1+\gamma^5\right)}{2} \frac{\mathrm{i}\left(\slashed p +\slashed k+M_{N_{i}}\right)}{\left(p+k\right)^2-M_{N_{i}}^2+\mathrm{i} \varepsilon} Y_{\mu i}\frac{\left(1-\gamma^5\right)}{2} u_{\mu}\left(p\right) \nonumber \\
& + & \frac{\mathrm{i}}{k^2-m_{\eta^{ \pm}}^2+\mathrm{i} \varepsilon} \frac{\mathrm{i}}{(k+q)^2-m_{\eta^{ \pm}}^2+\mathrm{i} \varepsilon} (-\mathrm{i} e)\varepsilon_\mu\left(2 k^\mu+q^\mu\right)
\end{eqnarray}

\begin{eqnarray}
    & = & - \frac{\varepsilon_\mu e}{4} \sum_{i} Y_{e i}^{*} Y_{\mu i}\int_{\mathbb{R}^4} \frac{\mathrm{d}^4 k}{(2 \pi)^4} \frac{N^{\mu}}{ABC} 
\end{eqnarray}

Using the Feynman parameters, we can combine the denominators to get
\begin{equation} 
 \frac{1}{ABC}  = 
\frac{1}{(\left(p+k\right)^2-M_{N_{i}}^2+\mathrm{i} \varepsilon)} \times \frac{1}{(k^2-m_{\eta^{ \pm}}^2+\mathrm{i} \varepsilon)} \times \frac{1}{((k+q)^2-m_{\eta^{ \pm}}^2+\mathrm{i} \varepsilon)} = 2 ! \int \frac{d \alpha_1 d \alpha_2 \theta\left(1-\alpha_1-\alpha_2\right)}{D^3} 
\end{equation} 
where
\begin{equation}
\begin{aligned}
D & =\alpha_1\left[(p+k)^2-M_{N_{i}}^2+\mathrm{i} \varepsilon)\right]+\alpha_2\left[((k+q)^2-m_{\eta^{ \pm}}^2+\mathrm{i} \varepsilon)\right]+\left(1-\alpha_1-\alpha_2\right)\left[(k^2-m_{\eta^{ \pm}}^2+\mathrm{i} \varepsilon)\right] \\
& =\left(k+\alpha_1 p+\alpha_2 q\right)^2-a^2 +\mathrm{i} \varepsilon
\end{aligned}
\end{equation}
with
$$
a^2=\alpha_1 M_{N_{i}}^2+\left(1-\alpha_1\right) m_{\eta^{ \pm}}^2
$$
To simplify the integral, we can shift the integration variable, $k \rightarrow k-\alpha_{1} p -\alpha_{2} q$. 

\begin{equation}
 \frac{1}{ABC} = 2 ! \int \frac{d \alpha_1 d \alpha_2 \theta\left(1-\alpha_1-\alpha_2\right)}{[k^2-a^2 +\mathrm{i} \varepsilon]^3}
\end{equation}
\begin{equation}
    N^{\mu} = \bar{u_{e}}\left(p-q\right) \left(2 k^\mu-2\alpha_{1} p^\mu+(1-2\alpha_{2})q^\mu\right)\left(1+\gamma^5\right)\left(\slashed k+(1-\alpha_{1})\slashed p -\alpha_{2} \slashed q+M_{N_{i}}\right)\left(1-\gamma^5\right)u_{\mu}\left(p\right)
\end{equation}

\begin{equation}
    N^{\mu} = \bar{u_{e}}\left(p-q\right) 2\left(1+\gamma^5\right)\left(-2\alpha_{1}(1-\alpha_{1}) p^\mu m_{\mu} \right)u_{\mu}\left(p\right) + \cdots \hspace{1cm} (\slashed p u_{\mu}\left(p\right) = m_{\mu} u_{\mu}\left(p\right))
\end{equation}

Under the shift($k \rightarrow k-\alpha_{1} p -\alpha_{2} q$), we get for the $p.\varepsilon$ term, which contributes to $\mu \rightarrow e \gamma$.

\begin{eqnarray}
\mathrm{i} \mathcal{M} & = & (p.\varepsilon)[\bar{u_{e}}\left(p-q\right) \left(1+\gamma^5\right)u_{\mu}\left(p\right) ]2 m_{\mu} e \sum_{i} Y_{e i}^{*} Y_{\mu i}  \int d \alpha_1 d \alpha_2 \theta\left(1-\alpha_1-\alpha_2\right) \alpha_{1}(1-\alpha_{1}) \nonumber \\
& + & \int_{\mathbb{R}^4} \frac{\mathrm{d}^4 k}{(2 \pi)^4} \frac{1}{[k^2-a^2 +\mathrm{i} \varepsilon]^3 }  
\end{eqnarray}

Momentum integration yields
\begin{equation}
\int_{\mathbb{R}^4} \frac{\mathrm{d}^4 k}{(2 \pi)^4} \frac{1}{[k^2-a^2 +\mathrm{i} \varepsilon]^3 } = \frac{\mathrm{i}}{32 \pi^2 a^2 }
\end{equation}
\begin{equation}
 \mathrm{i} \mathcal{M} =  2 \mathrm{i}(p.\varepsilon)[\bar{u_{e}}\left(p-q\right) \left(1+\gamma^5\right)u_{\mu}\left(p\right) ]\frac{m_{\mu} e}{32 \pi^2} \sum_{i} Y_{e i}^{*} Y_{\mu i}  \int d \alpha_1 d \alpha_2 \theta\left(1-\alpha_1-\alpha_2\right) \frac{\alpha_{1}(1-\alpha_{1})}{a^2}   
\end{equation}

The contribution to the invariant amplitude $A$ is then

\begin{equation}
 A(\mu \rightarrow e \gamma) =  \frac{m_{\mu} e}{16 \pi^2} \sum_{i} \frac{Y_{e i}^{*} Y_{\mu i}}{m_{\eta^{ \pm}}^2}  \int_{0}^{1} d \alpha_1 \frac{\alpha_{1}(1-\alpha_{1})^2}{2[\left(1-\alpha_1\right)+\alpha_1 \frac{M_{N_{i}}^2}{m_{\eta^{ \pm}}^2}] }   
\end{equation}
\begin{equation}
   \int_{0}^{1} d \alpha_1 \frac{\alpha_{1}(1-\alpha_{1})^2}{2[\left(1-\alpha_1\right)+\alpha_1 x] } = \frac{1-6x+3x^2+2x^3-6x^2\log x}{12 (1-x)^4} = j(x)
\end{equation}
\begin{equation}
 A(\mu \rightarrow e \gamma) =  \frac{m_{\mu} e}{16 \pi^2} \sum_{i} \frac{Y_{e i}^{*} Y_{\mu i}}{m_{\eta^{ \pm}}^2}  j\left(\frac{M_{N_{i}}^2}{m_{\eta^{ \pm}}^2}\right)  
\end{equation}
Decay width
\begin{equation}
   \Gamma(\mu \rightarrow e \gamma) = \frac{m_{\mu}^3}{16 \pi} |A|^2  = \frac{m_{\mu}^3}{16 \pi} \frac{m_{\mu}^2 e^2}{(16\pi^2)^2} \left|\sum_{i} \frac{Y_{e i}^{*} Y_{\mu i}}{m_{\eta^{ \pm}}^2}  j\left(\frac{M_{N_{i}}^2}{m_{\eta^{ \pm}}^2}\right)\right|^2
\end{equation}
\begin{equation}
   \Gamma(\mu \rightarrow e \gamma) = \frac{m_{\mu}^5 \alpha_{em}}{1024 \pi^4} \left|\sum_{i} \frac{Y_{e i}^{*} Y_{\mu i}}{m_{\eta^{ \pm}}^2}  j\left(\frac{M_{N_{i}}^2}{m_{\eta^{ \pm}}^2}\right)\right|^2 \hspace{0.6cm} (\alpha_{em} = \frac{e^2}{4\pi})
\end{equation}
\begin{equation}
       \Gamma(\mu \rightarrow e \nu \Bar{\nu}) = \frac{m_{\mu}^5 G_{F}^2}{192 \pi^3} 
\end{equation}
\begin{equation}
    \frac{BR(\mu \rightarrow e \gamma)}{BR(\mu \rightarrow e \nu \Bar{\nu})} = \frac{\Gamma(\mu \rightarrow e \gamma)}{\Gamma(\mu \rightarrow e \nu \Bar{\nu})} = \frac{m_{\mu}^5 \alpha_{em}}{1024 \pi^4} \times \frac{192 \pi^3}{m_{\mu}^5 G_{F}^2} \left|\sum_{i} \frac{Y_{e i}^{*} Y_{\mu i}}{m_{\eta^{ \pm}}^2}  j\left(\frac{M_{N_{i}}^2}{m_{\eta^{ \pm}}^2}\right)\right|^2
\end{equation}
\begin{equation}
    BR(\mu \rightarrow e \gamma) = BR(\mu \rightarrow e \nu \Bar{\nu}) \times \frac{3 \alpha_{em}}{16 \pi G_{F}^2}\left|\sum_{i} \frac{Y_{e i}^{*} Y_{\mu i}}{m_{\eta^{ \pm}}^2}  j\left(\frac{M_{N_{i}}^2}{m_{\eta^{ \pm}}^2}\right)\right|^2
\end{equation}

\FloatBarrier

\section{Annihilation, Production and Detection of DM}
\label{sec:appendix}

\begin{figure}[th]
        \includegraphics[height=20cm]{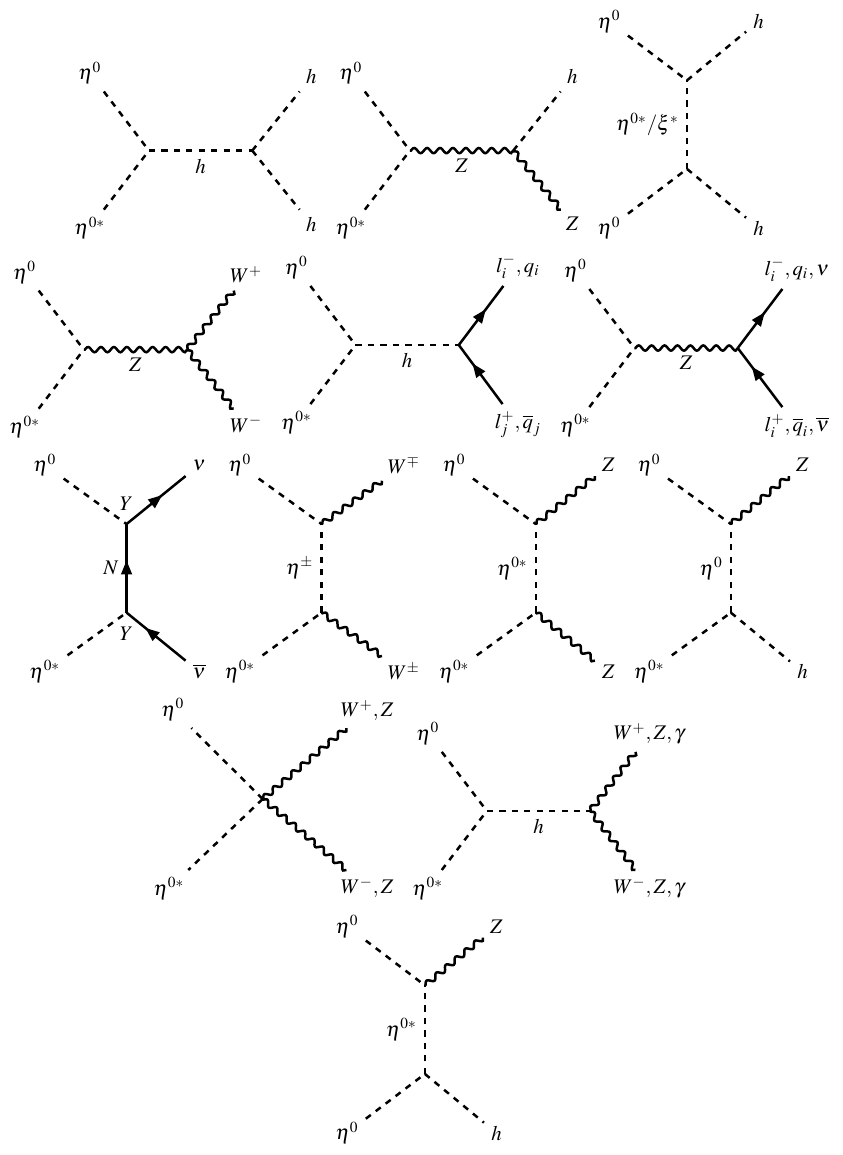}
        \caption{Relevant diagrams (Annihilation Channels) for computing the relic density of $\eta^{0}$ dominated DM candidate.}
                \label{fig:anihieta}
\end{figure}

\begin{figure}[th]
        \includegraphics[height=20cm]{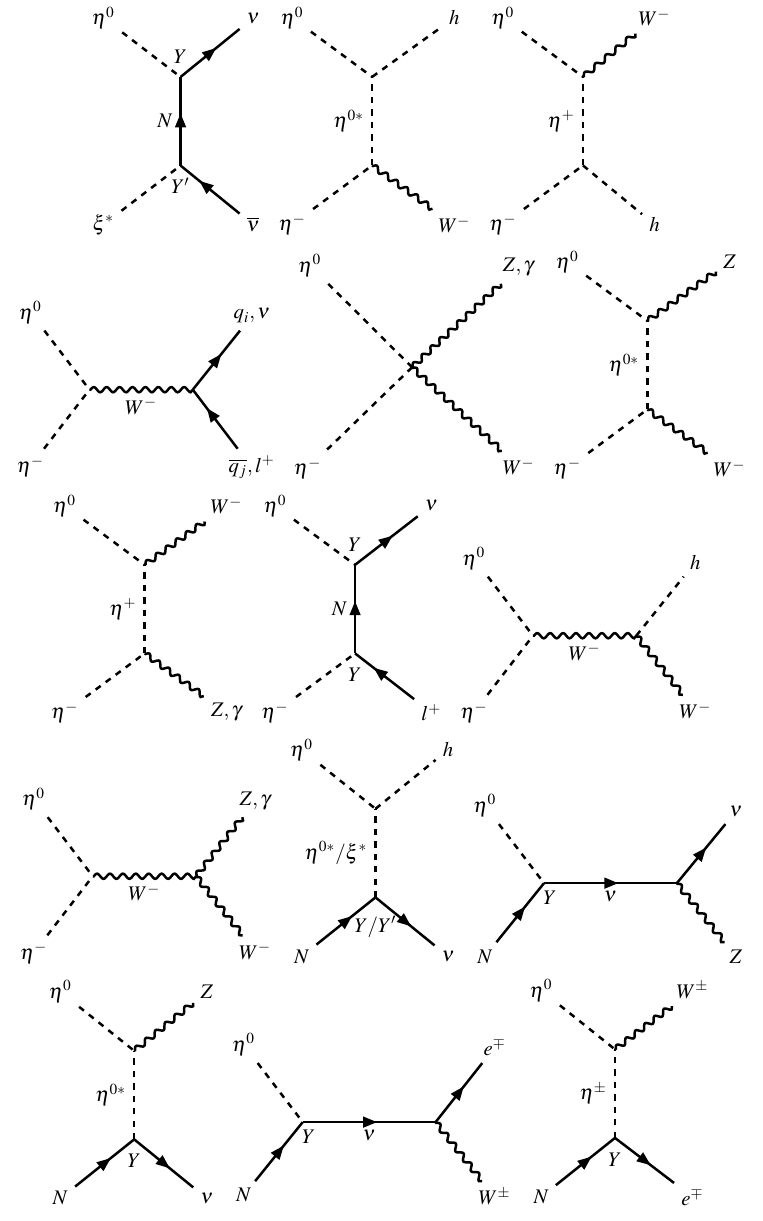}
        \caption{Relevant diagrams (Co-annihilation Channels) for computing the relic density of $\eta^{0}$ dominated DM candidate.}
                \label{fig:coanihieta}
\end{figure}

\begin{figure}[th]
        \includegraphics[height=8cm]{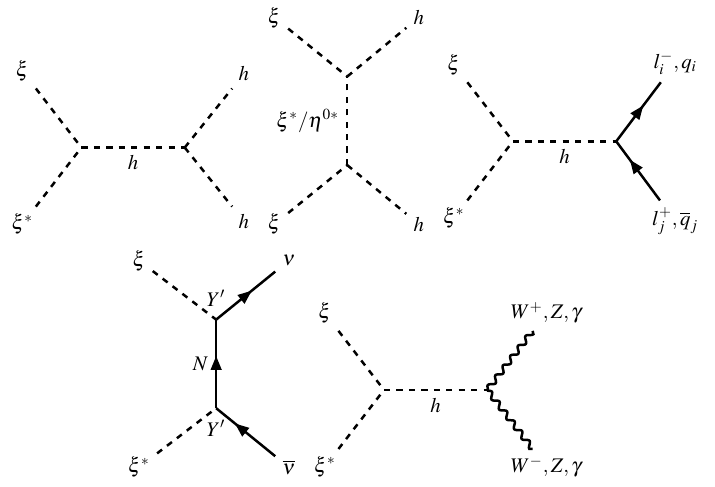}
        \caption{Relevant diagrams (Annihilation Channels) for computing the relic density of $\xi$ dominated DM candidate.}
                \label{fig:anihixi}
\end{figure}

\begin{figure}[th]
        \includegraphics[height=4cm]{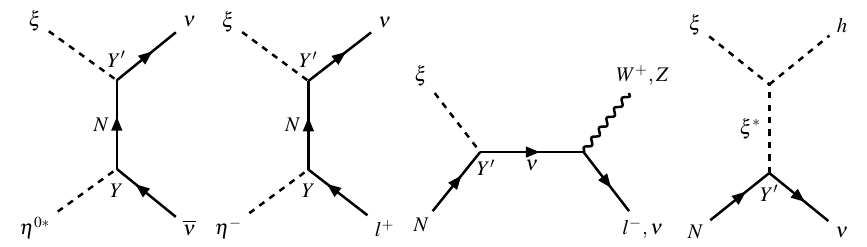}
        \caption{Relevant diagrams (Co-annihilation Channels) for computing the relic density of $\xi$ dominated DM candidate.}
                \label{fig:coanihixi}
\end{figure}

\begin{figure}[th]
        \includegraphics[height=4cm]{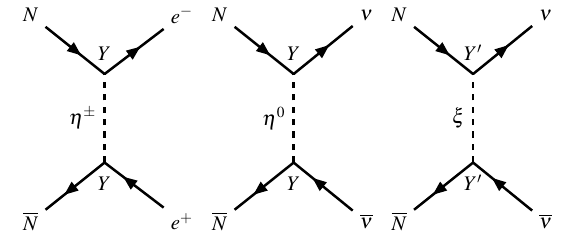}
        \caption{Relevant diagrams (Annihilation Channels) for computing the relic density of fermion DM candidate.}
                \label{fig:anihiN}
\end{figure}

\begin{figure}[th]
        \includegraphics[height=13.5cm]{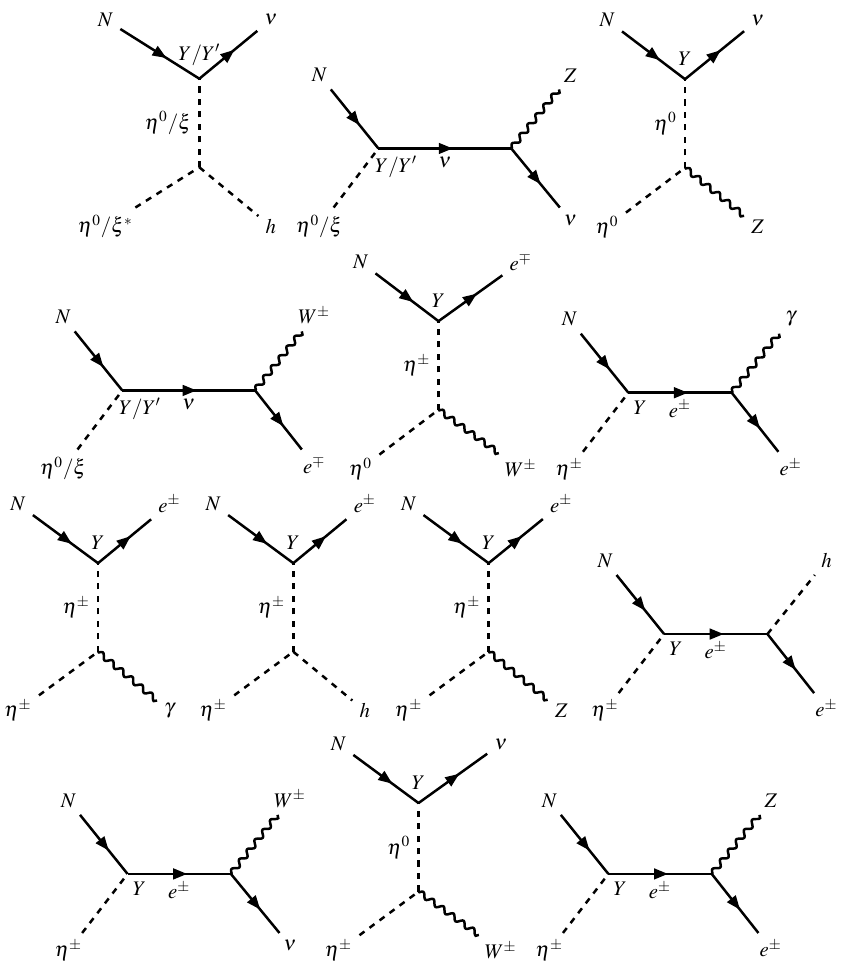}
        \caption{Relevant diagrams (Co-annihilation Channels) for computing the relic density of fermion DM candidate.}
                \label{fig:coanihiN}
\end{figure}

\begin{figure}[th]
        \includegraphics[height=5cm]{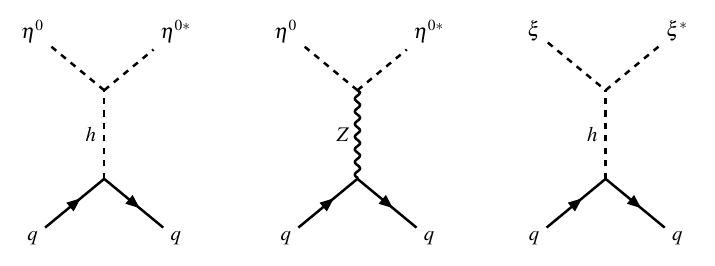}
        \caption{Relevant diagrams for the direct detection of the scalar DM candidate.}
                \label{fig:ddetaxi} 
\end{figure}

In Figs.~\ref{fig:anihieta} to \ref{fig:coanihiN}, we list the possible diagrams for production/annihilation of DM, relevant in the early universe, for the cases in which the DM is mainly a doublet scalar, singlet scalar or a fermion, respectively. In  Fig. \ref{fig:ddetaxi} we show the direct detection prospects of the scalar DM by exchange of a Higgs or $Z$ bosons.

\FloatBarrier

\section*{Acknowledgements}
Authors would like to acknowledge the SARAH-4.14.5 \cite{Staub:2015kfa} and SPheno-4.0.5 \cite{Porod:2011nf} to compute all vertex diagrams, mass matrices, and tadpole equations. Additionally, the thermal contribution to the dark matter relic abundance, as well as the cross sections for dark matter-nucleon scattering, are evaluated using micrOMEGAS-5.3.41 \cite{Belanger:2014vza}. SY would like to acknowledge the funding support by the CSIR SRF-NET fellowship. 

\bibliography{bibliography.bib,ref.bib}
\bibliographystyle{utphys}
\end{document}